\documentclass[a4paper,fleqn]{cas-dc}

\usepackage[authoryear,longnamesfirst]{natbib}
\usepackage{algorithm}
\usepackage{latexsym}
\usepackage{graphicx}
\usepackage{amsfonts}
\usepackage{amsmath}
\usepackage{tabularx}
\usepackage{textcomp}
\usepackage{fontawesome}
\usepackage{tabularx}
\usepackage{wasysym}
\usepackage{subfig}
\usepackage{algpseudocode}
\usepackage{changes}
\usepackage{makecell}
\usepackage{adjustbox}
\usepackage{svg}
\usepackage{textcomp}
\usepackage{stfloats}
\usepackage{verbatim}
\usepackage{threeparttable}
\usepackage{listings}
\usepackage{url}

\def\tsc#1{\csdef{#1}{\textsc{\lowercase{#1}}\xspace}}
\tsc{WGM}
\tsc{QE}

\begin{document}

\let\WriteBookmarks\relax
\def\floatpagepagefraction{1}
\def\textpagefraction{.001}

\def \sn {\textsc{Actminer}}

\shorttitle{M.Ma et.al / Knowledge-Based Systems}

\shortauthors{<short author list for running head>}

\title [mode = title]{\sn{}: Applying Causality Tracking and Increment Aligning for Graph-based Threat Hunting}

\author[1]{Mingjun~Ma}



\author[1]{Tiantian~Zhu}



\author[1]{Shuang Li}
\author[1,2]{Tieming~Chen}
\author[1]{Mingqi Lv}
\author[3]{Zhengqiu~Weng}
\fnmark[*]
\cortext[1]{Corresponding author}
\ead{derisweng@wzut.edu.cn}
\author[4]{Guolang~Chen}
\fnmark[*]
\ead{2019f098@zju.edu.cn}

\affiliation[1]{organization={Zhejiang University of Technology},
country={China}}
\affiliation[2]{organization={Zhejiang Key Laboratory of Visual Information Intelligent Processing},
country={China}}
\affiliation[3]{organization={Wenzhou University of Technology},
country={China}}
\affiliation[4]{organization={Wenzhou Polytechnic},
country={China}}

\credit{<Credit authorship details>}


\maketitle
\begin{abstract}
To defend against advanced persistent threats on the endpoint, threat hunting employs security knowledge, such as cyber threat intelligence (CTI) , to continuously analyze system audit logs through retrospective scanning, querying, or pattern matching, aiming to uncover attack patterns/graphs that traditional detection methods (e.g., recognition for point of interest) fail to capture. However, existing threat hunting systems based on provenance graphs face challenges of high false negatives (FNs), high false positives (FPs), and low efficiency when confronted with diverse attack tactics and voluminous audit logs. To address these issues, we propose a system called \sn{}, which constructs query graphs from descriptive relationships in CTI reports for precise threat hunting (i.e., graph alignment) on provenance graphs. First, we present a heuristic search strategy based on equivalent semantic transfer to reduce FNs. Second, we establish a filtering mechanism based on causal relationships of attack behaviors to mitigate FPs. Finally, we design a tree structure to incrementally update the alignment results, significantly improving hunting efficiency. Evaluation on the DARPA Engagement dataset demonstrates that compared with the SOTA POIROT, \sn{} reduces FPs by 39.1\%, eliminates all FNs, and effectively counters adversarial attacks.
\end{abstract}

\section{Introduction}\label{sec:intro}
Advanced persistent threats (APTs) aim to infiltrate specific institutions to obtain critical asset information and sensitive data, posing immense threats and impacts. Data provenance \cite{bates2015trustworthy} is widely used to analyze basic events (e.g., a sensitive file written by a malicious process) in a step-by-step manner to detect and investigate APT attacks on hosts. Nowadays, researchers have increasingly begun to employ provenance graphs in the field of APT attack detection on hosts.
\par
Existing provenance-based systems for detecting APTs fall into the following two main categories: rule-based \cite{milajerdi2019poirot,milajerdi2019holmes,hossain2020combating,xie2018pagoda,bishop2005introduction,kruegel2004intrusion,xiong2020conan,zhu2023aptshield,hossain2017sleuth} and learning-based \cite{altinisik2023provg,zengy2022shadewatcher,wang2022threatrace,han2020unicorn,manzoor2016fast,rehman2024flash,yang2023prographer} systems. The policies presented by rule-based systems are difficult to sustain in a constantly changing system environment, analysts must frequently update the rule base to adapt to new types of attacks, and the lag in defense results in frequent occurrences of false negatives (FNs). From the perspective of detection granularity, learning-based detection methods can be categorized into graph-level \cite{han2020unicorn,manzoor2016fast,altinisik2023provg,yang2023prographer} and node/edge-level \cite{zengy2022shadewatcher,wang2022threatrace,rehman2024flash} detection. Graph-level detection typically involves learning the characteristics of benign graphs and using certain techniques, such as clustering, to discern abnormal ones. However, in an anomalous subgraph \cite{yang2023prographer,han2020unicorn,manzoor2016fast}, not all nodes/edges are necessarily associated with the attack. Thus, experts still need further analysis to accurately pinpoint the malicious attack path. In contrast, node/edge-level detection methods can obtain the point of interest (POI), making them more direct and effective than graph-level detection methods in locating the attack candidate. However, alerts that solely focus on nodes and edges do not fully reveal the panorama of the attack, and analysts still need to spend a significant amount of time evaluating whether the generated POIs are false positives (FPs). According to CrowdStrike’s 2024 Global Threat Report \cite{CrowdStrike2024}, the time to compromise a host has gone from 84 min in 2022 to 62 min in 2023. This means that if an attack is not detected and responded to in a timely manner, the attacker will likely perform lateral movement \cite{lateral-movement}, which will cause more hazards. 
\par
Considering the shortcomings of the above detection systems, POIROT \cite{milajerdi2019poirot} has proposed a threat hunting approach based on graph alignment. This approach extracts the query graph from cyber threat intelligence (CTI) reports and then performs graph matching on the provenance graph to capture malicious behaviors. Due to the inherent interpretability of the query graph, its matching results can reflect comprehensive information about the attack, enabling analysts to respond accurately and promptly to the attack.
\par
However, POIROT still faces the following three limitations:
\par
- \textbf{Semantic Gap (C1)}. How to apply the attack knowledge extracted from CTI reports to solve the problems of attack camouflage and persistence. The extracted query graphs are often difficult to be directly mapped one-to-one to the provenance graph (e.g., files with the same type but different names). These mismatches between the two graphs can lead to inaccurate detection outcomes.
\par
- \textbf{Temporal and Causality Missing (C2)}. How to detect causal relationships of attacks in dynamic scenarios to minimize erroneous alerts. Solely concentrating on a single behavior can overwhelm the hunting system with an influx of alerts, thereby hindering its ability to effectively identify and mitigate truly malicious activities (e.g., discerning the semantic differences in the access to sensitive files, such as $/etc/passwd$, between normal processes and malicious processes requires consideration of causal relationships). This disregard may lead to more imprecise detection, rendering the system ineffective in countering sophisticated threats.
\par
- \textbf{Data Explosion Dilemmas (C3)}. How to minimize memory overhead and enhance the efficiency of threat hunting. Existing solutions assume an ideal scenario for datasets, that is, researchers assume that a complete attack can be discovered within a single batch of data. However, APT exhibits persistence, and attack chains may span across different batches of data (e.g., data from the first and third days). Repeated scans on ever-expanding datasets introduce significant overhead.
\par
In this work, we propose \sn{}, a threat hunting system that combines causality tracking and incremental aligning to efficiently and accurately dig attack chains. \textbf{To tackle C1}, \sn{} constructs a heuristic search strategy based on equivalent semantic transfer (EST) to counter phenomena such as attack camouflage, persistence, and evasion. We fuse the data information of inter-entity interactions through entities and their contextual semantics to accurately capture malicious behaviors. \textbf{To address C2}, \sn{} constructs a filtering mechanism based on the causal relationships of attack behaviors, and ignores unreasonable entity context relationships. \sn{} employs the causal motivation behind attacks to guide threat hunting, ensuring the interpretability of hunting results and minimizing FPs. In other words, we provide a more accurate hunting result by excluding unreasonable (attack-irrelevant) paths based on the causal relationship through temporal sequences. \textbf{To deal with C3}, \sn{} constructs a tree structure to incrementally update the alignment results, thereby avoiding the significant overhead caused by rescanning multiple batches of redundant data.
\par
We evaluate the effectiveness and efficiency of \sn{} on the dataset provided by DARPA-TC program \cite{DARPA3, darpa-tc}. Our results reveal that \sn{} surpasses existing provenance-based threat hunting systems in terms of detection precision and recall. Moreover, \sn{} reduces the computational overhead and eliminates redundant searches. By deploying \sn{}, security analysts can effectively analyze attack chains and formulate countermeasures, significantly alleviating the workload. In summary, the main contributions of our work are as follows:
\begin{itemize}
    \item Unlike traditional attack detection methods, we propose a provenance-based threat hunting system, i.e., \sn{}, to accurately capture attack chains.
    \item We introduce a heuristic search strategy based on EST and a filtering mechanism based on causal relationships of attack behaviors to ensure the precision and recall of \sn{}. 
    \item We propose a tree structure to incrementally update the alignment results, effectively addressing persistent APT attacks and the continuous growth of graph data. 
    \item We comprehensively evaluate our system, POIROT~\cite{milajerdi2019poirot}, and MEGR-APT \cite{MEGR-APT} on the dataset from the DARPA-TC program. The results demonstrate the efficiency of \sn{} in capturing the attack chain, highlighting its resistance to adversarial attacks.
\end{itemize}

\section{Background Knowledge}\label{sec:appendix_background}
\subsection{Advanced Persistent Threats and Threat Hunting}\label{Sec:apt}
Advanced Persistent Threats (APTs) are sophisticated cyberattacks characterized by their covert nature, targeted focus, and emphasis on data extraction. APTs are meticulously planned and involve multiple phases aimed at infiltrating and maintaining persistence in the target system undetected. APT actors select tools tailored to the target environment and the objectives of their mission \cite{milajerdi2019holmes,xiong2020conan}.
\par
The capabilities of APT actors have evolved over time as these groups expand their target scope, thereby adopting enhanced or new Tactics, Techniques, and Procedures (TTPs). These groups can develop tailored malware and data exfiltration methods. APT threat modeling is an extension of general threat modeling. It recognizes that attacks or campaigns are typically conducted through a series of phases. The Cyber Kill Chain \cite{lockheedmartinckc} and MITRE's ATT\&CK framework \cite{MITREATTCK} were developed based on extensive observations of APTs and related threat intelligence. 
\par
Threat hunting is a proactive cybersecurity approach that capitalizes on the distinct characteristics of APTs to detect and mitigate these threats. APTs follow a multi-phase attack lifecycle, as outlined in models like the Cyber Kill Chain and MITRE ATT\&CK framework. Threat hunters analyze system logs, network traffic, and endpoint data to identify indicators of compromise (IOCs) or anomalous behaviors that match known APT TTPs. Recently, substantial research \cite{altinisik2023provg,MEGR-APT,milajerdi2019poirot,mei2021ctscopy,sun2023cyber} has demonstrated that by analyzing the characteristics of TTPs intelligence associated with APT attacks, it is possible to proactively prevent such attacks from occurring.
\subsection{Provenance Graph}\label{sec:2.1-Provenance Graph}
\par 
Provenance graphs possess potent semantic expressiveness and contextual association capabilities, embodying the concrete manifestation of kernel audit logs. They model all system entities within logs as nodes and the interactions between entities as edges, with both nodes and edges bearing attribute information. The nodes within the provenance graph are categorized into subjects and objects based on the direction of data movement. Edges represent the causal relationships between system entities, such as read/write file operations, execute executable file operations, create/clone process operations. By leveraging provenance graphs, security professionals can associate malicious entities with attack behaviors through causal analysis, revealing the complete picture of an attack \cite{berrada2020baseline}.
\subsection{CTI Report and Query Graph}\label{sec:CTI Report}
CTI reports~\cite{openIoc,stix,misp} encompass comprehensive information related to cyberattacks and attackers, with a particular emphasis on capturing detailed attack procedures, i.e., the intricate sequences of steps and techniques employed in multistage attacks. These reports provide in-depth representations of attack scenarios, potential impacts on target hosts, and the complex chains of causally- linked events that characterize APTs. Security professionals leverage CTI reports to formulate more targeted defense rules for preventing and identifying malicious attack behaviors. In recent years, substantial research \cite{altinisik2023provg,milajerdi2019poirot,gao2021enabling,husari2017ttpdrill,liao2016acing,lv2024trec} has demonstrated the successful application of CTI reports in threat detection and threat hunting. In this paper, we construct a directed graph, called the query graph, from the offensive and defensive knowledge \cite{satvat2021extractor}(attack entities and their causal relationships) extracted from CTI reports. Similar to provenance graphs, query graphs are directed graphs with attribute information.

\subsection{Graph Alignment}
\label{sec:2.3-Graph-Alignment}
\par
Graph alignment refers to the problem of detecting potential cyber intrusion behaviors by establishing an optimal subgraph mapping between a provenance graph ($G_{p}$) representing system activities across the entire system, and a query graph ($G_q$) representing attack pattern activities. The provenance graph $G_p$ = ($V_p$, $E_p$) consists of a node set $V_p$ representing system entities and events, and an edge set $E_p$. Meanwhile, the query graph $G_q$ = ($V_q$, $E_q$) comprises a node set $V_q$ representing attack patterns and an edge set $E_q$. The goal of graph alignment is to find a subgraph $G_m$ in $G_p$ that maximizes the matching degree with $G_q$:
\begin{center}
\begin{equation}
Gm = \operatorname*{argmax}_{G' \subseteq Gp}(M(G', Gq))
\end{equation}
\end{center}
Here, $M$ is the match function that calculates the matching degree between $G_q$ and a subgraph $G'$ of $G_p$. The best mapping from the attack query graph to the activity graph is obtained by solving this optimization problem, enabling the detection and tracking of cyber intrusion behaviors.

\section{RelatedWork}\label{sec:relatedwork}
\subsection{Provenance Graph-based Threat Hunting}
The provenance graph contains causal relationships between system events, and is a data structure that can be effectively utilized for cyber threat hunting. DeepHunter \cite{wei2021deephunter} utilizes neural tensor networks to judge the subgraph relationship through graph embedding. However, due to the need for multiple-to-multiple traversal comparisons between subgraphs, the efficiency of DeepHunter will decrease as the size of the provenance graph increases. ProvG-Searcher \cite{altinisik2023provg} establishes a coarse-grained threat hunting by employing a graph representation learning method on the subgraph, with the aim to enhance hunting efficiency. ThreatRaptor \cite{gao2021enabling} extracts structured threat behaviors from OSCTI and automatically synthesizes query statements to search for malicious activities. Unlike the above studies, \sn{}, inspired by POIROT \cite{milajerdi2019poirot}, tries to align attack graphs extracted from CTI reports with provenance graphs from system logs to find complete attack chain. However, \sn{} significantly reduces FPs, FNs, and system overhead through an optimized graph alignment algorithm.

\subsection{Graph Matching Algorithms}
Graph pattern matching refers to the problem of finding similarities between a small query graph and a large target graph, with widespread applications (e.g., database analysis, search engines, software plagiarism detection, and social networks). Based on whether the matching results are completely consistent, algorithms can be divided into two categories: graph isomorphism matching and graph approximate matching. The main idea of graph isomorphism matching algorithms is to iteratively map nodes from the query graph to the target graph one by one. Ullmann et al. \cite{ullmann1976algorithm} propose a backtracking-based algorithm that enumerates all subgraphs that satisfy the matching requirements using a depth-first search approach. However, the enumeration range expands as the scale of the graph gradually increases, resulting in low algorithm efficiency. To address this issue, Cordella et al. \cite{cordella2004sub} propose the VF2 algorithm, which improves algorithm efficiency by incorporating the verification order of query nodes. However, in practical applications, its time complexity is superlinear, and the need for secondary filtering consumes a significant amount of time. Shasha et al. \cite{shasha2002algorithmics} introduce the GraphGrep algorithm, which achieves fast matching by encoding node semantic information. Several other studies \cite{cheng2007fg,zhao2007graph,yan2004graph} use graph mining techniques to find subgraphs from databases and then employ filtering and optimization strategies to prune incorrect nodes. However, we cannot always assume that the nodes and edges of the query graph can be fully mapped to the target graph due to the diversity and complexity of APT techniques. Graph approximate matching algorithms often rely on heuristic methods to identify important nodes and then gradually expand to neighboring nodes \cite{milajerdi2019poirot,tian2007saga,tian2008tale,tong2007fast}. Tian et al. \cite{tian2007saga} use a graph distance model to measure the similarity between graphs. He et al. \cite{he2006closure} introduce an index-based algorithm to support subgraph queries and similarity queries. Several other works \cite{pohly2012hi,tong2007fast} consider the shape and edge attributes of the query graph. However, the aforementioned research overlooks adversarial knowledge. Milajerdi et al. \cite{milajerdi2019poirot} propose POIROT, which is similar to our work and utilizes node attributes and information flows between nodes for approximate matching. However, it may result in missed detections if an attacker intentionally takes a detour to achieve their goal. Furthermore, if the similarity score exceeds the threshold, POIROT stops hunting, and the obtained attack subgraph may not represent the optimal attack behavior. We develop a new matching technique to address these challenges.
\subsection{Incremental Graph Computation}
In real-world scenarios, graphs are typically large in scale and frequently updated over time. When graphs are updated, traditional batch processing methods require starting the computation from scratch, which is extremely time-consuming. In contrast to traditional batch processing algorithms, incremental matching only analyzes and matches the updated portion, utilizing previous matching results to maximize the reduction of redundant computations. Fan et al. propose the IncSIMMatch \cite{fan2013incremental} algorithm, which effectively reduces redundant computations by creating an index for the pattern graph. They then introduce an incremental computation method IncISO \cite{fan2017incremental}, 
where only the set of nodes within $d$ hops of the updated node $d$ in the data graph needs to be rematched as the affected region when nodes and edges change in the graph. Subsequently, researchers \cite{kao2016distributed,choudhury2015selectivity,kankanamge2017graphflow,idris2017dynamic,idris2020general,kim2018turboflux,min2021symmetric} introduce the idea of incremental computation and incorporated query search optimization strategies. For example, Sutanay et al. \cite{choudhury2015selectivity} construct the pattern graph as a binary tree and decompose it for storage in tree nodes, with data updates performed by searching leaf nodes. However, storing a large number of indexes consumes memory. Sun et al. \cite{sun2022depth} propose an exploration-based approximate matching technique; however, if the initially selected node is inappropriate, a large number of useless intermediate result values will be generated. \sn{} constructs a tree structure to incrementally update the alignment results and introduces a forgetting rate to maintain stable memory overhead.
\section{Motivation}\label{sec:motivation}
\subsection{Motivating Example}
\label{sec:2.4-Alert—Fatigue-solutions}
\par
Scenario: Consider the following scenario where an attacker exploits the feature of automatically executing login scripts (Reg.exe and \%temp\% \textbackslash art.bat\_2) during login initialization (mal) to establish persistence by adding the malicious script path to the registry (HKCU \textbackslash Environment\_R2). Subsequently, the attacker searches for network shares on the compromised computer to locate files and then collects sensitive data (/etc/passwd) from remote locations via shared network drives (host shared directories, network file servers, etc.). Finally, the data is transmitted over the network (162.66.239.75).
\par
\begin{figure*}
  \centering
  \includegraphics[width=\textwidth]{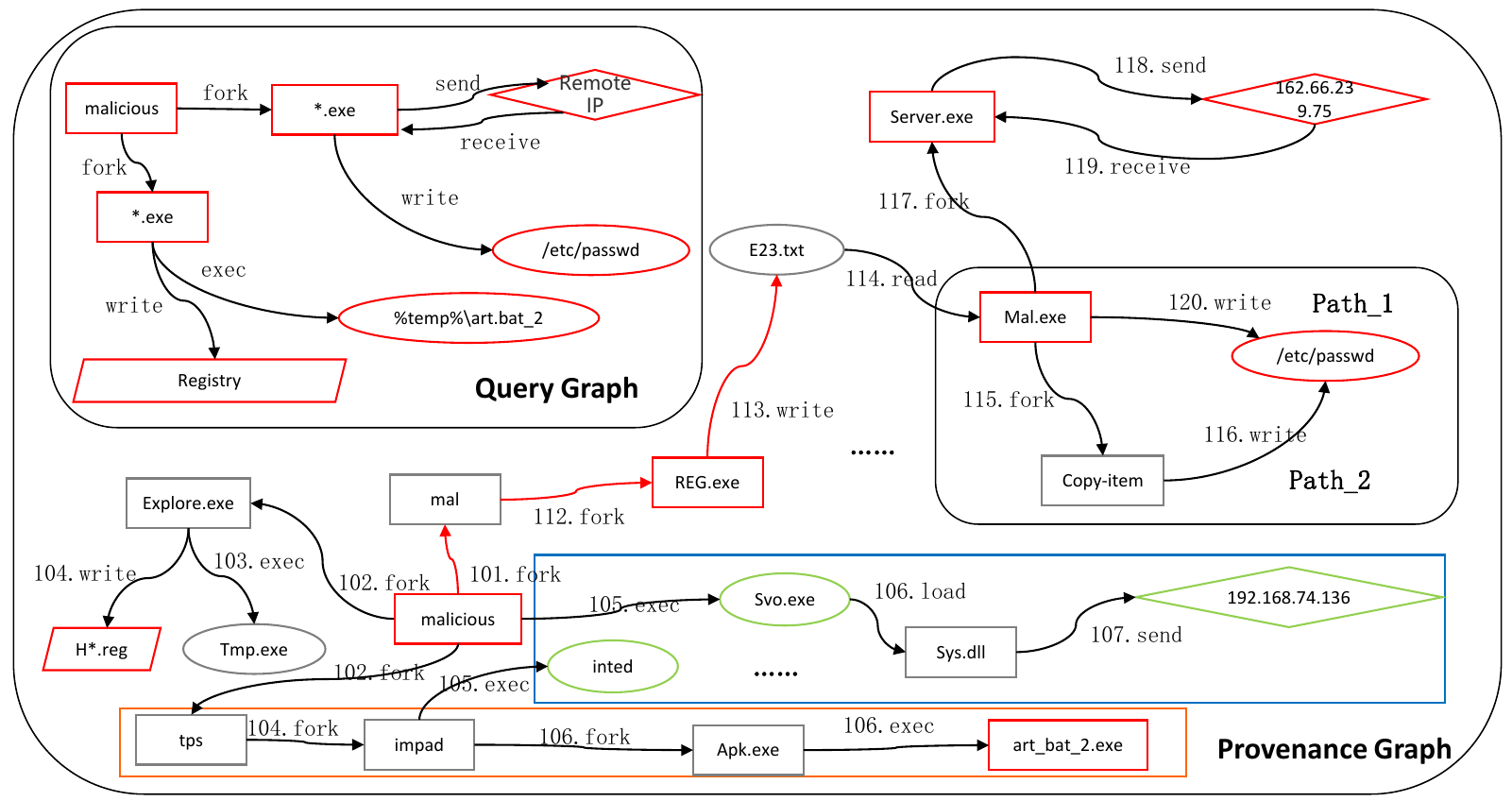}
    \caption{Motivating Example. The red nodes and edges depict the truly malicious behavior. In contrast, the blue outlines encompass false positive detection, where POIROT incorrectly identified benign system entities as malicious. The specific nodes are the points with green borders. Furthermore, the orange outlines highlight the instances of missed detection or false negatives, where POIROT failed to identify nodes that were indeed part of the attack chain.}
    \label{fig:Example}
\end{figure*}

As illustrated in Figure \ref{fig:Example}, this example includes two graphs: the top left depicts the attack query graph manually extracted from a CTI report, following the approach outlined in the POIROT.
\par
\noindent
\framebox{%
  \begin{minipage}[t]{\linewidth}
    The attack initiates by leveraging a malicious executable, malicious.exe, to obtain unauthorized system access. It then employs Registry modifications to establish persistence. Once a foothold is secured, the attacker can remotely issue commands and execute them on the system, executing an art.bat\_2 file in the temporary folder and facilitating actions such as exfiltrating sensitive data to an external IP, exemplified by the transfer of /etc/passwd containing user account information.
  \end{minipage}
}
This example traces a complete attack chain—from the initial execution of malicious.exe, through Registry-based persistence establishment, to the final remote retrieval of /etc/passwd.
\par
The right side depicts the provenance graph constructed from actual system logs capturing the observed execution behavior. The fragmented nature of attack scenarios, coupled with the constraint of limiting the hop count to existing threat hunting approaches, can lead to imprecise or incomplete results during the threat hunting process. In this paper, we transform the threat hunting problem into finding the attack query graph within the provenance graph.

Threat hunting methods centered around POIROT encounter several significant challenges: 
\par
\textbf{False Negative}. Semantic gaps exist between provenance graphs and attack query graphs due to the complexity of real enterprise environments. The manifestations of the same attack type may differ across systems, and attackers may utilize the same tools in various ways. For instance, entity names in the attack query graph may have varying representations in the underlying logs of different systems. In POIROT, regular expressions are employed to instantiate node names from the attack query graph for hunting searches in the provenance graph. However, if attackers modify their tactics, introducing technical variations, threat hunting systems struggle to detect different mutated attack samples (e.g., over 100 versions of the Carbanak malware were described in CTI reports). Furthermore, attackers can evade security detection through obfuscation, persistence, and evasion techniques. As illustrated, the attack query graph only describes data exfiltration over the network, whereas in the real environment, the attacker creates processes to read files, followed by another process reading those files, and finally transmitting them over the network. Overreliance on the simplistic threat hunting approach based on a predetermined number of hops may inadvertently overlook malicious activities that align with the intrinsic characteristics of attacks, ultimately resulting in detection effort failure. As shown in Figure \ref{fig:Example}, in a simulated scenario, as described in \cite{milajerdi2019poirot} Section 5, we set $C_{thr}$ to 3 but find that this limitation resulted in an incomplete capture of the attack chain. Consequently, it was unable to detect the art\_{bat} file. Moreover, with such hop count restrictions, attackers that are aware of the imposed limits could potentially evade hunting more easily across different scenarios, resulting in potential harm. The manual adjustment of the hop limit according to different scenarios poses significant challenges. Similarly, the same problem exists in other path-based detection \cite{hassan2019nodoze,hassan2020tactical,wang2020you} efforts.
\par
\textbf{False Positive.} Within real organizations, extensive legitimate user operations exhibit similarities with attack behaviors in log data. Normal behaviors may be misclassified as malicious if hunting rules are overly broad or incomplete. For example, when a user downloads network files through a browser, the browser collects user data and transmits it to its cloud server, while the downloaded network files may be flagged as “suspicious files” by the system, resembling malicious attack behaviors and triggering false alarms from the hunting system. Additionally, attackers may leverage tools/techniques to deceive hunting systems, which also leads to FPs. As highlighted in Figure~\ref{fig:Example}, a suspicious process \textit{Mal.exe} exhibits two paths for reading/writing sensitive files. According to the attack query graph, sensitive file access should occur before network transmission, while \textit{$path_1$} occurs after transmission. Therefore, \textit{$path_2$} represents the attacker’s actual operations in the environment. 

\textbf{High Overhead and Inefficiency.} Government and enterprise organizations typically need to collect data simultaneously from thousands of machines, easily amassing petabyte-scale data volumes. This massive data not only imposes substantial storage overhead but also significantly reduces hunting efficiency. Traditional hunting methods require offline storage and continuous back scanning of system log data, resulting in immense computational overhead for each hunting operation. Referring to Figure~\ref{fig:Example}, assume that all operations before node \textit{E23.txt} at time 114 have already occurred. When a security analyst attempts to hunt for threats solely based on the data collected after this time, the incremental data segment alone cannot effectively support the reconstruction of the complete attack chain represented by its query graph. When examining the issue holistically, the newly acquired data lacks the necessary evidence to capture the earlier stages of the multistep intrusion. Consequently, subsequent data collection would necessitate rescanning the previously available information, redundantly recomputing the provenance of data prior to a specific time frame. These redundant computations across multiple hunting activities introduce an unsustainable overhead, hindering the efficiency and scalability of the system and potentially allowing malicious activities to persist undetected for extended periods.

\section{System Design}\label{sec:sysdesign}
This section first introduces the overall architecture of the \sn{} system, followed by a detailed description of each module presented in \sn{}.

\subsection{System Overview}
\label{System:detail}
\textbf{Data Preparation Module (§~\ref{sec:3.2_preparation}).} The attack query graphs are extracted from threat intelligence reports, and provenance graphs are constructed based on extensive underlying logs. Duplicate events and orphan nodes within the provenance graphs undergo filtering, which is a necessary and common practice in existing work~\cite{inam2023sok}. 
\par
\textbf{Casual Relation and Semantic Processing Module (§~\ref{sec:3.2_casual}).} When a new attack query graph or provenance graph is generated, \sn{} will first categorize the entities into four classes. The provenance graph will then deliver to the next module while the attack query graph still needs to be processed. Next, \sn{} merges analogous actions in the attack query graph. Finally, \sn{} employs \textit{EST}, which traces potentially overlooked attack chains by tracking malicious semantics, to identify suspicious actions in the next module.
\par
\textbf{Threat Hunting and Incremental Aligning Module (§~\ref{sec:3.2_th}).}
\sn{} will hunt attack-related scenario by chaining suspicious semantic nodes and generating suspicious semantic tree. As time progresses, batch data is continuously inputted into the \sn{}, persistently updating our suspicious semantic trees and unveiling more latent malicious attacks. To control the memory consumption, \sn{} will store the unupdated tree branch to the database unless a certain behavior is observed related to this branch.
\par
\begin{figure*}[ht]
\centering
\includegraphics[width=1\linewidth]{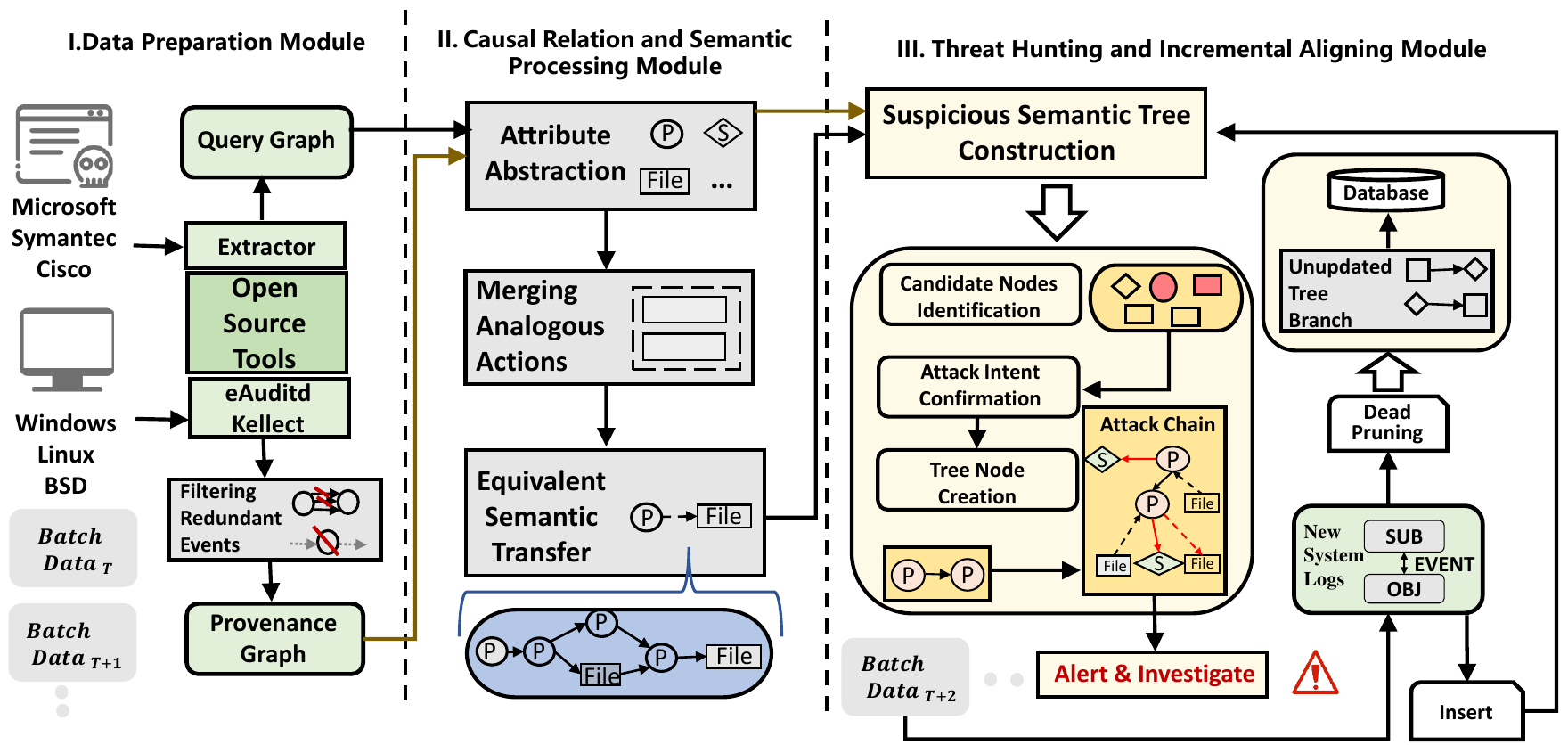}
\caption{The architecture of \sn{}, which consists of three core modules that synergistically facilitate comprehensive threat hunting and attack chain construction capabilities.  
}
\label{fig:System Architecture}
\end{figure*}

The basic architecture of \sn{} is shown in Figure~\ref{fig:System Architecture}, which can be divided into three modules: (\uppercase\expandafter{\romannumeral1}) the Data Preparation Module, (\uppercase\expandafter{\romannumeral2}) Casual Relation and Semantic Processing Module, and (\uppercase\expandafter{\romannumeral3}) Threat Hunting and Incremental Aligning Module. It is important to note that \sn{} continuously runs the above three modules as time progresses. Details of system design for each module are provided in Section~\ref{sec:3.2_preparation}, Section~\ref{sec:3.2_casual}, and Section~\ref{sec:3.2_th}, respectively. 

\par
\par
\subsection{Data Preparation Module}
\label{sec:3.2_preparation}
\par
This section describes the data preprocessing module for provenance graphs and query graphs. 
\subsubsection{Provenance Preparation}
Provenance graphs are composed of log data collected from various platforms by data collectors. In this work, we employ open-source tools, such as eAuditd~\cite{sekar2023eaudit}, Kellect~\cite{chen2022kellect}, and Event Tracing for Windows \cite{etw}, to gather relevant system logs from Linux and Windows environments. \sn{} transforms each event into a directed, time-stamped edge, in which the source node represents the object being acted upon. For any event $ et \in E$, \sn{} represents it as a quintuple $\langle{UID_{s}, UID_{o}, OP, Ti_{i}}\rangle$. $UID_s$ and $UID_o$ are unique identifiers for the subject and object of $et$, respectively. $OP$ denotes the type of $et$, and $Ti_i$ denotes the time when $et$ occurred.
\par
Directly processing such massive raw log data is extremely challenging. To address this, we perform pruning operations on the low-level log data. Specifically, redundant events without context are removed~\cite{zhu2021general, zhu2023aptshield}. This means that if the subject UID ($UID_s$), object UID ($UID_o$), and operation ($OP$) are identical and the timestamps ($T_i$) are consecutive, the most recent timestamp will be preserved. Furthermore, our methodology includes the removal of isolated nodes within the provenance graph. The isolated nodes refer to entities that lack any incoming or outgoing edges. For example, we find that the data contains many of these nodes, which manifest without any parent or child nodes. Moreover, none of the events contain a subject UID ($UID_s$) or object UID ($UID_o$) matching the UID of the node, lack contextual information, and fail to provide meaningful insights. Therefore, they must be removed without compromising the integrity of the graph representation.
\par
Concurrently, during the process of constructing provenance graphs, the node types are categorized based on the entity type contained within the logs. For instance, $F_{a}$ represents files involving user and system-sensitive information, such as the boot.ini file on Windows and /etc/passwd file on Linux. The specific categorization is summarized in Table \ref{table:category}.
\par
As presented in Table~\ref{table:category}, we are inspired by Aptshield~\cite{zhu2023aptshield} and Conan~\cite{xiong2020conan}. We then refined the classification heuristics and optimized the classification method we obtained from POIROT. Furthermore, we extensively gather CTI reports from various online sources and network channels, identifying specific file paths that exhibit heightened susceptibility to attacks. Consequently, we adapt the importance degree of these paths based on their frequency of occurrence to finally obtain ten distinct labels.
\begin{table}[h!t]
\caption{\centering A categorization of distinct entities and their corresponding label assignments.}
\resizebox{\linewidth}{!}{%
\label{table:category}
\begin{tabular}{|c|c|c|} 
\hline
\textbf{Entity} & \textbf{Tag} & \textbf{Description} \\ 
\hline
Process & P & Processes, threads spawned by system calls \\ 
\hline
\begin{tabular}[c]{@{}c@{}}User Configuration \\Sensitive Files\end{tabular} & Fa & \begin{tabular}[c]{@{}c@{}}Sensitive files containing user configuration \\information, such as /etc/passwd\end{tabular} \\ 
\hline
\begin{tabular}[c]{@{}c@{}}Application \\Configuration\\~Sensitive Files\end{tabular} & Fb & \begin{tabular}[c]{@{}c@{}}Sensitive files that contain configuration information \\about the application, such as /etc/mysql/my.cnf\end{tabular} \\ 
\hline
\begin{tabular}[c]{@{}c@{}}Log-sensitive\\~documents\end{tabular} & Fc & \begin{tabular}[c]{@{}c@{}}Sensitive files containing logging information, \\e.g. /etc/httpd/logs, e.g. etc/httpd/logs\end{tabular} \\ 
\hline
Library file & Fd & \begin{tabular}[c]{@{}c@{}}A collection of pre-compiled methods with extensions \\such as .lib, .a, .dll, .so, etc.\end{tabular} \\ 
\hline
Executable file & Fe & \begin{tabular}[c]{@{}c@{}}Files that can be loaded and executed by the operating system, \\with extensions such as: .exe, .vbs, etc.\end{tabular} \\ 
\hline
Temporary document & Ff & Temporary files generated by the system, e.g., /tmp/* \\ 
\hline
Other documents & Fg & \begin{tabular}[c]{@{}c@{}}A collection of other types of files, such as plain text files,\\~plain graph files, plain zip files, etc.\end{tabular} \\ 
\hline
Registration form & R & \begin{tabular}[c]{@{}c@{}}Unified management of hardware and software configurations,\\~including HKLM, HKCU, HKCR, HKCC, HKU, etc.\end{tabular} \\ 
\hline
Socket & S & \begin{tabular}[c]{@{}c@{}}Refers to a host on the Internet or \\a process in a host, e.g. 127.0.0.1\end{tabular} \\
\hline
\end{tabular}%
}
\end{table}
By analyzing CDM18 and CDM19, which refer to data definitions for DARPA’s E3 and E4 programs, respectively, we adopt events based on a few general fields in the CDM (i.e., events of read, write, fork, clone, create, execute, load, and inject), which were most commonly used in previous studies, such as Holme \cite{milajerdi2019holmes}, Sleuth \cite{hossain2017sleuth}, and MORSE \cite{hossain2020combating}. The details are presented in Table~\ref{event}.
\begin{table*}[]
\captionsetup{font={normalsize}}
\caption{Classification of different event types.}
\centering
\label{event}
\resizebox{\linewidth}{!}{%
\begin{tabular}{|c|l|l|l|l|l|}
\hline
\textbf{Number} & \multicolumn{1}{c|}{\textbf{EventType}} & \multicolumn{1}{c|}{\textbf{Subject}} & \multicolumn{1}{c|}{\textbf{Object}} & \multicolumn{1}{c|}{\textbf{Direction}} & \multicolumn{1}{c|}{\textbf{Description}} \\ \hline
1               & accept                                  & P                                     & S                                    & backward                                & Accepting socket connections                                  \\ \hline
2               & inject                                    & P                                     & P                                    & forward                                 & Running arbitrary code in the address space of independent activity processes.                                 \\ \hline
3               & clone                                   & P                                     & P                                    & forward                                 & Cloned subject                                     \\ \hline
4               & connect                                 & P                                     & S                                    & forward                                 & Connect to a socket type guest                              \\ \hline
5               & execute                                 & P                                     & F                                    & forward                                 & The subject invokes and executes the object                                 \\ \hline
6               & fork                                    & P                                     & P                                    & forward                                 & Creating the process body                                   \\ \hline
7               & load                                    & P                                     & F                                    & backward                                 & Load file to current workspace data                             \\ \hline
8               & write                                   & P                                     & F                                    & forward                                 & Open a file or directory (Object) and write information                        \\ \hline
9               & receive                                 & P                                     & S                                    & backward                                & Receive data through a connected socket port                           \\ \hline
10              & send                                    & P                                     & S                                    & forward                                 & Sending data through a connected socket port                           \\ \hline
11              & exit                                    & P                                     & P                                    & forward                                 & Process exit                                      \\ \hline
12              & unlink                                  & P                                     & F                                    & forward                                 & Deletion of individual files                                   \\ \hline
\end{tabular}
}
\end{table*}
\subsubsection{Query Graph Preparation.}
CTI reports describe attacks that have already occurred. We collect the latest threat intelligence from websites such as Microsoft and Symantec. Leveraging open-source tools such as Extractor \cite{satvat2021extractor}, we extract attack query graphs ($G_q$) from CTI reports. In our extraction process, we found that the graphs extracted by the Extractor exhibit issues, such as incompleteness and fragmentation. CRUcialG \cite{cheng2024crucialg} was then developed specifically to address and solve this issue. The graphs extracted by the Extractor are input into CRUcialG, and the graphs can represent the attack process and semantics by combining automation with expert experience.
\par
\begin{figure}
    \centering
    \includegraphics[width=1\linewidth]{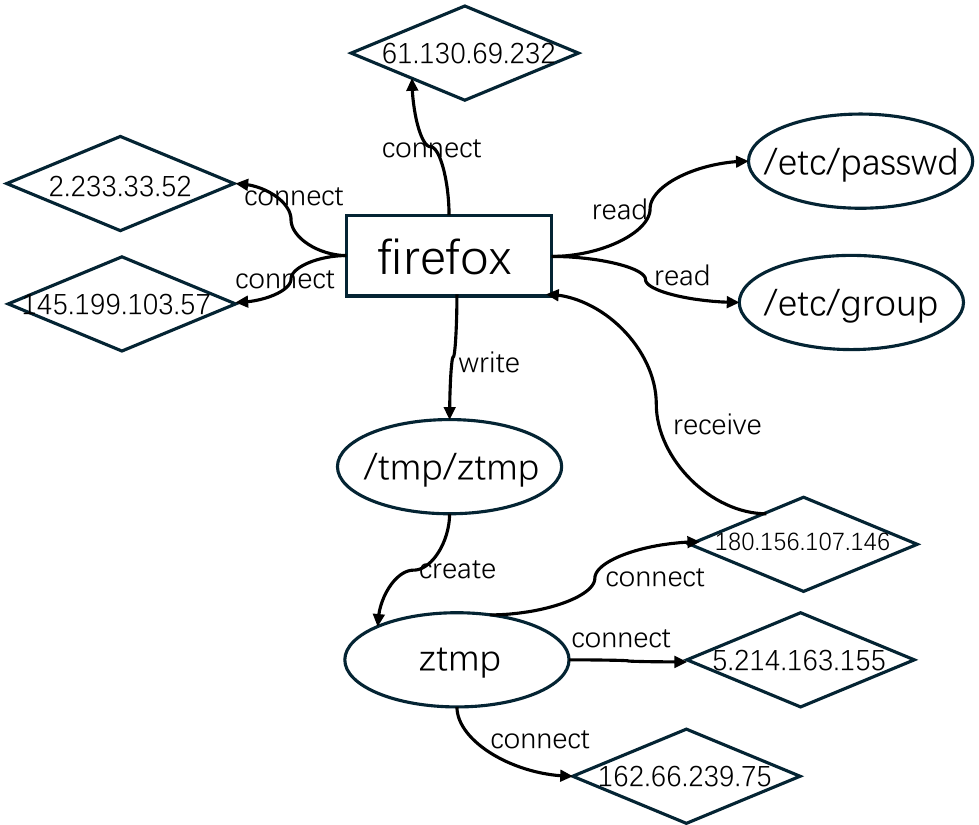}
    \caption{The type and details of query graph}
    \label{fig:ogq}
\end{figure}
\begin{figure}
    \centering    \includegraphics[width=0.85\linewidth]{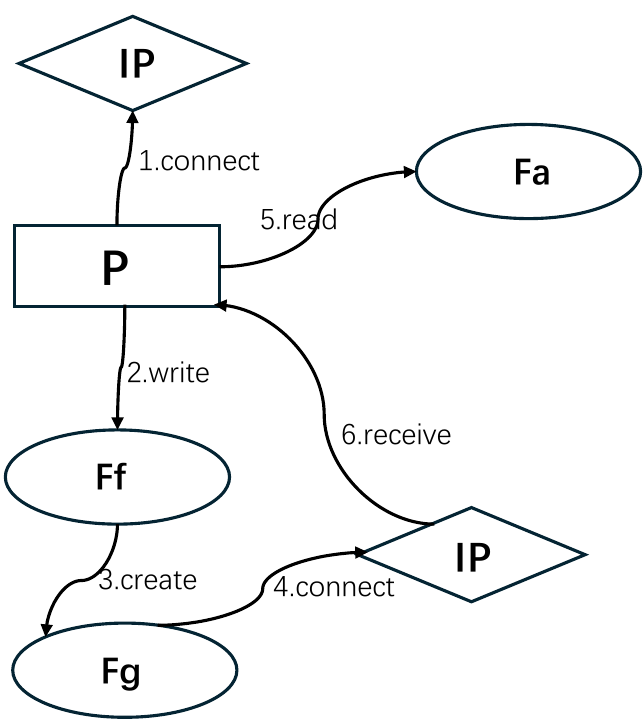}
    \caption{The type and details of the processed query graph}
    \label{fig:gq}
\end{figure}

We construct query graphs by analyzing CTI attack reports and abstracting attack patterns into graph representations. CTI reports are compiled by security analysts to document IOCs and malicious activities. Figure \ref{fig:ogq} illustrates an example query graph constructed from an attack report in the DARPA TC3 dataset \cite{darpa-tc}. Processes, files, and network flows are represented by rectangles, ovals, and diamonds, respectively. The attack initiates by compromising a Firefox browser extension, subsequently downloading a malicious executable file (“ztmp”) to the target host and establish initial access. The attack then executes this malicious file, creating a corresponding malicious process (“ztmp”) that connects to a command-and-control (C\&C) server via the external IP address 162.66.239.75. The malicious process conducts internal reconnaissance by scanning for open ports using “portscan” before deleting the malicious file to eliminate attack artifacts. Figure \ref{fig:gq} shows the query graph after processing.
\subsection{Casual Relation and Semantic Processing Module}
\label{sec:3.2_casual}
\par
In this section, we sequentially describe the process of Module II in Figure~\ref{fig:System Architecture}, i.e., merging analogous actions and EST. Briefly, analogous actions in attack query graphs are merged while employing an EST strategy. This process enhances attack hunting capabilities while simultaneously preparing the data for input to Module III.
\subsubsection{Merging Analogous Actions}
When \sn{} adopts the attack query graphs to perform the threat hunting tasks, it encounters the issue of \textit{rigid matching}, causing the difficulty to achieve alignment between attack query graphs and provenance graphs. Finally, it would result in high FN and FP rate in Module III.
\par
Additionally, the directly extracted $G_q$ is fixed, limiting its ability to generalize and effectively defend against variants of known attacks, as it is specifically tailored to documented attack patterns. A key insight of \sn{}, \textit{merging analogous actions}, is developed to address this challenge. This method integrates nodes with similar operations. For instance, instead of representing each individual file with a distinct node, the generalized $G_q$ may group files based on their types (e.g., $F_{a}, F_{b}, F_{c}$ in Table~\ref{table:category}) in the attack chain. Simultaneously, we consider the temporal relationships of attack events. Processes can be merged based on their functional attributes or the operations they perform, rather than being strictly tied to specific executable paths or process names. In essence, this method transforms the information flow from the source entity to the same target entity while preserving the semantic meaning of the source entity, so that equivalent events can be removed as redundant information, thereby achieving efficient compression of the data.
\subsubsection{Equivalent Semantic Transfer}
During the attack process, malicious processes controlled by the attacker interact with other entities~\cite{xiong2020conan}, causing malicious behaviors and effects to proliferate across the system through the intricate web of entity interactions and information propagation pathways and expanding the attacker’s control scope.
\par
\begin{figure}
    \centering
    \includegraphics[width=1\linewidth]{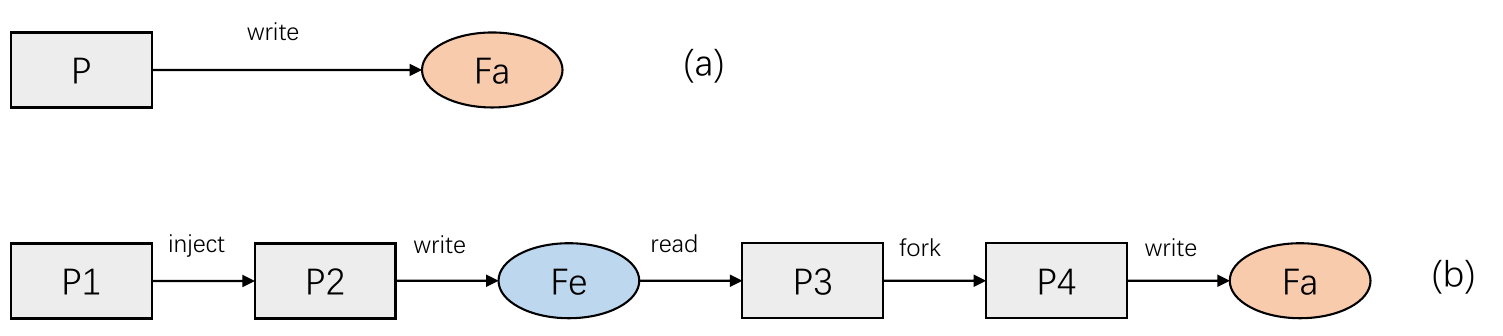}
    \caption{An Example of equivalent semantic transitivity.}
    \label{fig:EqS}
\end{figure}

Based on above finding, we construct the equivalent semantic transitivity strategy to address the FN issue. The key insight of this strategy is that the semantics of malicious behavior propagate with taking actions. Figure~\ref{fig:EqS}.(a) depicts a short attack path described in the query graph, where a process controlled by the attacker tampers with sensitive files. Figure \ref{fig:EqS}.(b) illustrates one of the attacker’s specific implementation approaches. First, a process P1 with suspicious semantics injects malicious code to create process P2, and then process P2 writes a suspicious file Fe. Next, another malicious process P3 accesses the tainted file. Subsequently, process P3 with suspicious semantics creates process P4. Finally, process P4 with suspicious semantics performs a write operation on sensitive files. Through this series of operations, the ultimate behavior of a suspicious process tampering with sensitive files Fa is achieved. Although the types of controlled entities change during the attack process, the ultimate goal remains unchanged. The operations in (b) are essentially the same as those in (a) and align with the attacker’s intent, making the two paths semantically equivalent.
\par
To comprehensively capture attack intents and mitigate attack evasion while addressing semantic gaps between $G_q$ and $G_p$, we integrated log data from the DARPA project, leveraged attack stage theories proposed in previous works such as Conan~\cite{xiong2020conan} and Aptshield~\cite{zhu2023aptshield}, and incorporated relevant descriptions from the CDM document \cite{DARPA3-CDM} to extract six equivalent semantic transitivity policies, as presented in Table \ref{table:rules}. Therefore, \sn{} can track and locate suspicious behavior on the host in real time by analyzing the data flow transmission, while systems like HOLMES~\cite{milajerdi2019holmes} cannot do this (according to Table 8 in HOLMES). In contrast, Sleuth~\cite{hossain2017sleuth} employs initial label propagation to detect suspicious behavior in specific steps. For regular processes, a malicious command line affects only the data trustworthiness tag (data t-tag), leaving the code trustworthiness tag (code t-tag) as “benign.” As a result, detection policies focused on the code t-tag do not trigger alerts based solely on the command line itself. Based on the context information of entities, these policies automatically determine whether an event is attack related or not. The subject represents a process, the object represents different types of entities connected with the subject, and the direction indicates the information flow between the subject and the object. For example, the third policy represents process $\rightarrow$ file: if the process in the write event is considered as attack related, suspicious semantics will propagate from the process to the file. If another process reads the file containing suspicious semantics, the suspicious semantics will further propagate from the file to that process. We then use a heuristic search algorithm integrating equivalent semantic transitivity in Section \ref{sec:3.2_th}.

\begin{table}
\centering
\caption{Equivalent Semantic Transitivity Policies in the Context of Generalized Attack Pattern Identification and Matching.}
\label{table:rules}
\resizebox{\linewidth}{!}{
\begin{tabular}{llll} 
\hline
\multicolumn{1}{c}{\textbf{Subject}} & \multicolumn{1}{c}{\textbf{Object}} & \multicolumn{1}{c}{\textbf{Direction}} & \multicolumn{1}{c}{\textbf{Requisites}} \\ 
\hline
P & P & forward & \begin{tabular}[c]{@{}l@{}}$\exists$p.semantics$\epsilon$\{SuspiciousLabel\}\\ $\wedge$[Event\_Fork(p, p') \textbar{} Event\_Create(p, p')\textbar{} \\Event\_Clone(p, p')]: p'.semantics.add("SSuspiciousLabel")\end{tabular} \\
P & P & forward & \begin{tabular}[c]{@{}l@{}}$\exists$p.semantics$\epsilon$\{SuspiciousLabel\}\\ $\wedge$Event\_Inject(p, p'): p'.semantics.add("SuspiciousLabel")\end{tabular} \\
P & F & forward & \begin{tabular}[c]{@{}l@{}}$\exists$p.semantics$\epsilon$\{SuspiciousLabel\}\\ $\wedge$Event\_Write(p, f): f.semantics.add("SuspiciousLabel")\end{tabular} \\
P & F & backward & \begin{tabular}[c]{@{}l@{}}$\exists$f.semantics$\epsilon$\{SuspiciousLabel\}$\wedge$f.tag$\epsilon$\{Fd, Fe\} \\ $\wedge$Event\_Execute(p, f)\textbar{}Event\_Load(p, f): p.semantics.add("SuspiciousLabel")\end{tabular} \\
P & F & backward & \begin{tabular}[c]{@{}l@{}}$\exists$f.semantics$\epsilon$\{SuspiciousLabel\}\\ $\wedge$Event\_Read(p, f): p.semantics.add("SupiciousLabel")\end{tabular} \\
\hline
\end{tabular}
}
\end{table}

\subsection{Threat Hunting and Incremental Aligning Module}
\label{sec:3.2_th}
\subsubsection{Suspicious Semantic Tree Construction}
\label{sec:sstc}
An event contains the interaction information between entities and can be transformed into an information flow, which can be further classified into data flows and control flows. Data flows indicate dependencies in data content, reflecting the data propagation path (e.g., a process reading a file), while control flows primarily refer to process creation relationships (e.g., a parent process creating a child process). In the threat hunting module, data flows and control flows will be jointly abstracted into a suspicious semantic tree. The following shows the process of generating a suspicious semantic tree, as presented in algorithm~\ref{hunting}:
\par
\textbf{Step 1: Finding Candidate Nodes.} To capture malicious behaviors in the provenance graph constructed from low-level system logs that match the patterns in the corresponding query graph, our system first searches for all nodes in the provenance graph with attributes identical to those of the entity node in the query graph. These candidate nodes are collected into a list, referred to as the candidate set $\textit{FC(i)}$, which is associated with the query node (line 7).
\par
\textbf{Step 2: Confirming Attack Intent.} As the query graph $G_q$ carries clear temporal features and causal relationships, we leverage these information to guide the attack detection reconstruction and reconstruction processes. This enables us to quickly determine the initial intrusion location, relevant entity nodes, and sequence of attack events. Such a query graph can assist analysts in searching malicious behaviors effectively. The $\textit{ExtractRelevantNodes}$ function is subsequently invoked to identify the critical nodes within the query graph that are essential for comprehending the attack methodology (line 8). This function operates by determining the next potential action nodes in $G_p$ based on the preceding step $F_c$ in $G_q$.
The $\textit{ReconstructAttackSequence}$ algorithm reconstructs the attack sequence (line 9), carefully aligning with the temporal and causal patterns within $G_q$. Unlike indiscriminate search methods, the approach is strategically guided by specific target nodes, accurately capturing the attacker’s intended progression. The function determines the matching order by meticulously tracing the sequence of attacks delineated in the graph. \sn{} evaluates whether nodes introduce malicious semantics, which is identified by first fixing the target nodes and then analyzing behavior within $G_q$. This process involves tracing actions from the fixed nodes to determine patterns indicating malicious intent, forming the basis for further analysis.
\par
To optimize the traversal and analysis process, the algorithm stores context information for each node (lines 11–15). This step avoids redundant traversals of the graph and enables the algorithm to order the edges based on timestamps, resulting in an ordered hunting sequence (lines 16–19). This sequence is instrumental in guiding the subsequent steps of the algorithm. It will calculate the reciprocal of the length of the shortest path between nodes in Gq and Gp as the path score. For each node in Gq, the candidate node with the highest contribution value will be selected as the fixed node.
\par
\textbf{Step 3: Creating Tree Nodes.} The final step involves the creation of tree nodes (lines 19–27). Given the disparity in size between the query graph and the provenance graph, the $\textit{CreateTreeNode}$ function (line 31) is used to map each query graph to one or more subgraphs in the provenance graph that exhibit similar patterns. The function specifically creates a branch in the tree for the attack entry point. This mapping is facilitated by the hunting sequence obtained in the previous step, enabling efficient detection navigation against the large and complex dataset.
\par
After the tree nodes are created, when the attack progression exceeds the total attack sequence, the system raises an alert to notify security analysts of this anomalous situation.
\begin{algorithm}
\caption{Threat Hunting Algorithm}
\label{hunting}
\begin{algorithmic}[1]
     \Require $G_{q}$, $G_{p}$ 
     \Ensure Suspicious Subgraph $G_s$\\
     /*\textbf{DataProcessing}*/
     \State $G_{q} \gets MergeSimilarEntity(G_{q})$
     \State $L_{Se} \gets GetQuerySequences(G_{q})$
     \State $f \gets GetSetQueryGraphFlow(G_{q})$
     \State $G_{p} \gets MergeSimilarEvent(G_{p})$\\
     /*\textbf{ThreatHunting} and \textbf{Incremental Aligning}*/
     \State $FC \gets FindCandidateNodes(G_q, G_p)$
    \State $relevant\_nodes \gets ExtractRelevantNodes(G_{q})$
    \State $attack\_sequence \gets ReconstructAttackSequence(G_{q})$
    \State $hunting\_sequence \gets \emptyset$
    \For{$n \in relevant\_nodes$}
        \State $context \gets RetrieveNodeContext(n)$
        \State $hunting\_sequence.add(context)$
    \EndFor
    \State $hunting\_sequence.sort(key=lambda x: x.timestamp)$
    \State $visited \gets \emptyset$
    \State $fixed\_candidates \gets \emptyset$
\State $seqNum \gets 0$
\For{$context \in hunting\_sequence$}
    \State $q\_node \gets context.corresponding\_query\_node$
    \If{$q\_node \notin visited$}
        \State $candidates \gets FC[q\_node]$
        \State $fixed\_candidates[q\_node] \gets candidates$
        \State $visited[q\_node] \gets True$
    \Else
        \State $candidates \gets fixed\_candidates[q\_node]$
    \EndIf
    \For{$c \in candidates$}
        \If{$IsMaliciousSemantic(c)$}
            \State $seqNum \gets seqNum + 1$
            \State $node \gets CreateTreeNode(c, seqNum)$
            \State $G_s.add(node)$
        \EndIf
    \EndFor
\EndFor
\State \textbf{return} $G_s$
\end{algorithmic}
\end{algorithm}
\par
\begin{figure*}[]
\centering
\includegraphics[scale=0.55]{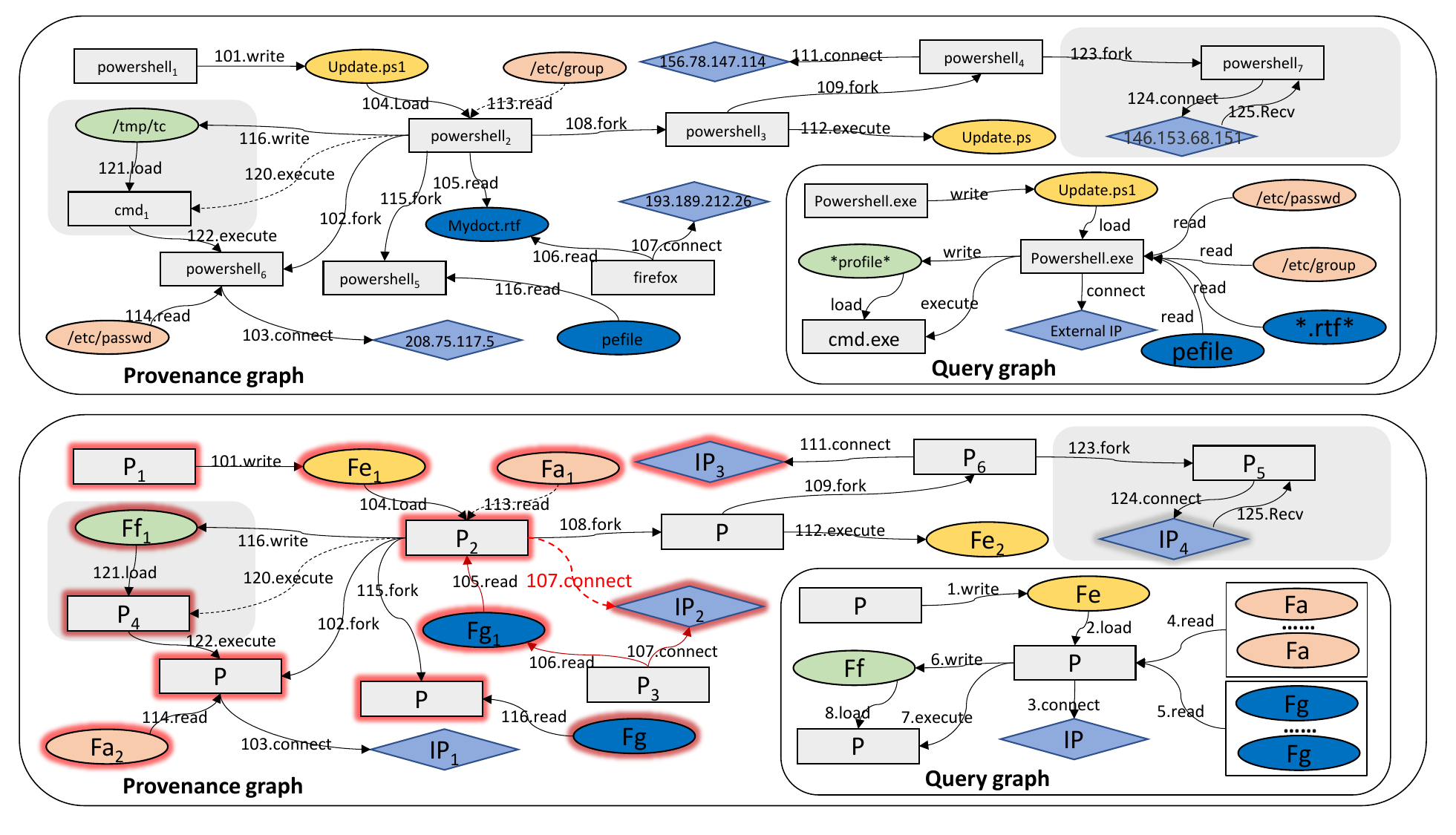}
\setlength{\abovecaptionskip}{0.1cm}
\captionsetup{font={normalsize}}
\caption{A case study of strategies such as POIROT and the underlying scenarios that our \sn{} faces in the same scenario. Where the dashed part indicates that the middle contains the multi-step behaviour whether black or red color, and the red line and red box part indicate the real captured entities and events. }
\label{fig:Explain}
\end{figure*}

We present an example for better comprehension to illustrate the aforementioned methodology. We aim to find suspicious subgraphs similar to the query graph in the provenance graph shown in Figure \ref{fig:Explain}. The representation above illustrates a concrete instantiation of the graph, whereas the depiction below presents an abstraction of the model. Assuming we start from process P1 within the red box as the starting point of the attack chain (corresponding to the above is $powershell_1.exe$), according to the query graph, the next step should be to find an executable file associated with P1, with the edge semantics being a write event. In the graph, only the Fe1 node (corresponding to the above is $update.ps1$) satisfies this condition; therefore, we can generate a tree node that stores a variety of data, including the unique identifier of the query graph, event type, and relevant temporal information to assist subsequent hunting tasks. This step is largely analogous for both approaches, with only marginal differences in storage efficiency and temporal performance. The detail is provided in Section~\ref{section:4.4}.

Similarly, $P2$ will also be identified and generate a tree node (corresponding to the above is $powershell_2.exe$). However, multiple similar scenarios may arise when the flows in $G_q$ come to be associated with the socket. In the $G_p$, the path from P2 to the socket IP2 (indicated by the red arrow) satisfies the previously defined equivalent semantic transitivity policy, indicating that P2 and P3 share the same semantic information. Therefore, IP2 can be retained as a suspicious node while generating a tree node and preserving the temporal relationship from P3 to IP2. However, for the above scenario, with the absence of strategy and rigid aligning rules, a FP result is produced. Likewise, the path from P2 to IP3 also satisfies the equivalent semantic transitivity policy (it should be noted that intermediate nodes will be represented in the form of equivalent semantic attributes within the properties of the initial P1 node). However, although the path from P2 to IP1 also represents P$\rightarrow$IP with the edge semantics of a connect operation, it occurs before the previous node (Fe1$\rightarrow$P2 at time 104), violating the sequence of the attack. Consequently, this path is ignored. Fa can be found as the same.
\par
Next, we need to find P$\rightarrow$Ff (with the edge semantics of a write operation); however, there are no files of the Ff type in the current provenance graph (assuming the current time has not yet reached 116), so the system needs to wait for new data to arrive.

\subsubsection{Incremental Aligning}
As time progresses, the log data generated by hosts will continue to increase. For any newly added logs after a period of time, traditional threat hunting systems need to rescan the entire dataset with a larger volume. To address the inefficiency of traditional threat hunting systems in analyzing incremental streaming data, we adopt an incremental graph computation method to hunt for attacks and update suspicious semantic trees. First, we search for new candidate nodes in the newly arrived provenance graph based on the node attributes in the query graph. Next, we divide the impact of new data on the suspicious subtrees into two parts, i.e., new data that affects the existing suspicious semantic subtrees and new data that is unrelated to the existing suspicious subtrees. Furthermore, to manage memory overhead effectively, we implement a forgetting rate to reduce memory consumption. We transfer the nodes that remain unupdated for a period of 6 h (this duration can be adjusted according to different circumstances) into the database and create a corresponding index for them. The index includes the attributes of the node itself and its parent node, enabling rapid localization of relevant nodes in the event of a subsequent occurrence. This allows for the efficient retrieval and reinstatement of these nodes into memory.
\par
For the former, through the affected tree nodes, we can obtain the current attack progress and their mapped nodes in the provenance graph and query graph. Through the sequence of suspicious candidate nodes, we can then determine the next suspicious entity to hunt. Finally, we judge whether the suspicious states are met in the candidate node set of the suspicious entity. If this is achieved, we then construct a new tree node.
\par
For the latter, we first determine whether the node has candidate nodes. The absence of candidate nodes means that this part of the data does not have an entry point for attacks, indicating that this part of the data is considered benign. However, if there are candidate nodes, we will start rebuilding the suspicious subtree from its candidate nodes.
\par
As an example, the shaded part in Figure \ref{fig:Explain} represents the newly added data. For the new data, we determine whether it affects the existing results. The query graph ($G_q$) awaits the arrival of a pattern where process P2 writes file Ff1. If so, a new node representing Ff1 is added to the graph. According to the query graph, this P$\rightarrow$Ff1 (write) is the desired one-hop attack path, for the above graph is $powershell.exe$ $\rightarrow$profile (write). In the query graph, P$\leftarrow$Fg (read) occurs before P$\rightarrow$Ff (write), since the previous operation reads an ordinary file with weak attack relevance. Therefore, if the provenance graph does not contain the corresponding related connected flow and nodes, it indicates that the operation did not introduce new suspicious attack semantics and the next attack target should be further explored. We then search for the target corresponding to P$\rightarrow$P (execute). There exists an execute action in the P2 process node to another process P4 (the dashed line indicates no direct connection between nodes, and there are multiple hops), generating new tree nodes corresponding to P2. Similarly, interactions with sockets are present in the newly added data on the right shaded part. As described above, the semantics represented by P6 are the same as those of P2, and the P2 node can be mapped to the P node in the query graph P$\rightarrow$IP (connect), so P6$\rightarrow$IP4 (connect) is equivalent to P2$\rightarrow$IP4 (connect), allowing the generation of a new tree node.
\section{Evaluation}\label{sec:evaluation}
Our evaluation aims to answer the following five questions:
\begin{itemize}
    \item \textbf{RQ1}: How effectively can \sn{} detect the attacks especially in terms of false alarms?
    \item \textbf{RQ2}: How robust is \sn{} against adversarial attacks?
    \item \textbf{RQ3}: How important are the components we design for assisting threat hunting?
    \item \textbf{RQ4}: How efficient is \sn{} compared with SOTA in terms of runtime overhead?
    \item \textbf{RQ5}: How robust is \sn{} in a benign dataset?
\end{itemize}
\textbf{DataSet.} \sn{} is evaluated using three datasets: DARPAE3 \cite{DARPA3}, DARPAE4, and Simulated Environments. DARPAE3 is an open-source dataset, whereas the DARPAE4 dataset is not publicly available.
Theia and FiveDirections are both from DARPA Engagement 3, while Trace is from DARPA Engagements 3 and 4. The data of Theia and Trace was collected from Linux, while that of FiveDirections was collected from Windows 7. As presented in Table \ref{tab:summaryExperDatasets}, the duration encompasses both benign activities and attack-related activities within a dataset, wherein only a small portion of the total time frame involves actual attacks. Detailed description of both actual attack behaviors and benign operations is presented in Table~\ref{table:attack tell}.
\par
As presented in Table \ref{table:attack tell}, these attacks include the following scenarios: malicious file downloads and execution, information collection and exfiltration, Firefox memory attacks, backdoor extensions, and phishing emails. Furthermore, we employ Kellect \cite{chen2022kellect} to extract benign data from the Windows platform, while we use SPADE \cite{gehani2012spade} to acquire benign log data from Linux. All benign data includes various daily system operations, such as network browsing and file operations. We also use the OpTC dataset which contains benign activities of 500 Windows hosts over 7 days. These benign datasets contain billions of audit records on Windows, Linux, and FreeBSD. 
\begin{table}[t]
\centering
\caption{The detail of the attack and benign datasets.}
\label{table:attack tell}
\resizebox{\linewidth}{!}{%
\begin{tabular}{|cllllll|clllllll|}
\hline
\multicolumn{7}{|c|}{Scenario}             & \multicolumn{8}{c|}{Behavior}                                  \\ \hline
\multicolumn{7}{|c|}{E4-Trace case1}   & \multicolumn{8}{c|}{Malicious file download and execute}       \\ \hline
\multicolumn{7}{|c|}{E4-Trace case2}   & \multicolumn{8}{c|}{Information gather and exfiltration}       \\ \hline
\multicolumn{7}{|c|}{E4-Trace case3}   & \multicolumn{8}{c|}{Malicious file download and sensitive file exfiltration}       \\ \hline
\multicolumn{7}{|c|}{E4-Trace case4}   & \multicolumn{8}{c|}{In-memory attack with firefox}             \\ \hline
\multicolumn{7}{|c|}{E3-FiveDir case1} & \multicolumn{8}{c|}{Pine backdoor}                             \\ \hline
\multicolumn{7}{|c|}{E3-FiveDir case2} & \multicolumn{8}{c|}{Phishing E-mail Link with macro viruses}   \\ \hline
\multicolumn{7}{|c|}{E3-Trace case1} & \multicolumn{8}{c|}{Firefox backdoor and load malicious software}                      \\ \hline
\multicolumn{7}{|c|}{E3-Trace case2} & \multicolumn{8}{c|}{Firefox backdoor and deploy malicious programme}                   \\ \hline
\multicolumn{7}{|c|}{E3-Trace case3}   & \multicolumn{8}{c|}{Phishing E-mail}                           \\ \hline
\multicolumn{7}{|c|}{E3-Theia case1}   & \multicolumn{8}{c|}{Firefox Backdoor and privilege escalation} \\ \hline
\multicolumn{7}{|c|}{Win Benign Data}    & \multicolumn{8}{c|}{Account operation, network communication and application activity} \\ \hline
\multicolumn{7}{|c|}{Linux Benign Data}  & \multicolumn{8}{c|}{User Login, application operation and network interaction}         \\ \hline
\end{tabular}%
}
\end{table}

\begin{table}[]
\centering
\caption{The summary of the experimental dataset. Column 1 specifies the name of dataset, and Column 2 denotes the corresponding duration. Columns 3 and 4 indicate the number of nodes and edges, respectively. Column 5 represents the number of attack nodes.}
\label{tab:summaryExperDatasets}
\resizebox{\linewidth}{!}{%
\begin{tabular}{ccccc}
\hline
\textbf{Datasets}  & \textbf{Duration Time} & \textbf{\# N} & \textbf{\# E} & \textbf{\% of Attack Nodes}  \\ 
\hline
E3-Trace          & 310h                   & 1.950M        & 9.053M        &  37890                   \\
E3-FiveDirections & 210h                   & 1.287M        & 2.577M        & 2956                    \\
E3-THEIA          & 168h                   & 960.357K      & 2.352M        & 14781                    \\
E4-Trace          & 8h                   & 3.035M        & 13.586M       & 39582                     \\
Benign Linux      & 240h                   & 2.385M        & 3.891M        & 0                     \\
Benign Windows    & 192h                   & 5.324M        & 12.856M       & 0                     \\ 
\hline
\textbf{Avg}      & 234.667h               & 2.490M        & 7.386M        & 0.669\%                     \\
\hline
\end{tabular}%
}
\end{table}

\par
\textbf{Detector for Comparison.} We first evaluate the performance of \sn{} by comparing it with the following state-of-the-art threat hunting methods, including both graph alignment–based and machine learning–based approaches:
\begin{itemize}
    \item \textbf{MEGR-APT}~\cite{MEGR-APT}: It proposes a graph neural network–based attack representation learning approach to effectively detect suspicious subgraphs that match attack query patterns. It extracts comprehensive embeddings for both query and provenance graphs by taking node types as initial features and then utilizes a neural tensor network for graph alignment.
    \item \textbf{ProvG-Searcher}~\cite{altinisik2023provg}: It introduces a graph simplification strategy that preserves diverse behaviors in provenance graphs, enhancing learning and search capabilities. It leverages order embeddings for efficient evaluation of subgraph relationships in the embedding space.    \item\textbf{POIROT}~\cite{milajerdi2019poirot}: It advances APT attack detection by modeling threat hunting as an inexact graph pattern matching problem that leverages CTI. Through similarity-based alignment between these graph representations, POIROT enables effective detection of suspicious activities that match known attack patterns, even when the match is not exact.
\end{itemize}
\par
Since POIROT is not open source, we have attempted to implement the steps outlined in the POIROT paper using the algorithm and used point-level ground truth to evaluate POIROT and \sn{}. Based on MORSE \cite{hossain2020combating}, we conduct manual annotation of the DARPAE4 dataset. For the DARPAE3 dataset, we acquired the GT labels from existing studies: Karios \cite{cheng2023kairospracticalintrusiondetection}, Flash \cite{rehman2024flash}, ThreaTrace \cite{wang2022threatrace}, and REAPr \cite{reapr-ground-truth}. Due to the unavailability of the query graphs manually extracted by the authors in POIROT, for fairness, our query graph is uniformly extracted by the Extractor \cite{satvat2021extractor} in our experimental setting. However, during the extraction process, we encounter graph disconnections and incomplete attack chains, among other problems. Therefore, we use the state-of-the-art method CRUcialG \cite{cheng2024crucialg} to assist in obtaining the query graph.
\subsection{RQ1: How effectively can \sn{} detect the attacks especially in terms of false alarms?}
\label{sec:4.1}
Table~\ref{table:eva_metric} presents the performance of \sn{} and POIROT on our evaluation datasets. \sn{} consistently surpasses POIROT, achieving superior precision, recall values, and lower number of FN/FP. Compared with POIROT, \sn{} utilizes entity attributes and action semantics to generate a more generalizable query graph. This provides brief abstracted entity information to take a graph alignment, subsequently reducing FNs and enhancing precision and recall. As the POIROT paper lacks evaluation on E4 dataset, we execute POIROT on E4 to obtain evaluation results. The findings demonstrate that \sn{} significantly outperforms POIROT, as E4 attacks are more challenging to detect due to well-blended malicious activity. POIROT relies solely on name and simple types as the query graph’s features, which is insufficient. In contrast, the entity attributes and action semantics employed by \sn{} offer more semantic information for each node and are less sensitive for nodes with certain types of characteristics, making it harder for attack nodes to conceal themselves.
\par
At first sight, \sn{} shows incremental improvement compared with POIROT in terms of FN. This is attributed to \sn{} considering the correlation between the semantics of multihop nodes. Table~\ref{table:eva_metric} summarizes the performance of \sn{} and POIROT on all datasets. \sn{} does not miss any malicious nodes, i.e., its average FNs are 0, reduced by 61 compared to POIROT. On average, the FP nodes generated by \sn{} ($\sim$ 389 nodes) is 1.91 $\times$ less than that of POIROT. \sn{} demonstrates a notable improvement in precision over POIROT, achieving a 2.96\% higher precision score while also outperforming POIROT in terms of recall with a 1.94\% increase.
\par
We reimplemented MEGR-APT and ProvG-Searcher and conducted a comparative evaluation. As presented in Table \ref{tab:megr-apt}, \sn{} consistently outperforms both MEGR-APT \cite{MEGR-APT} and ProvG-Searcher \cite{altinisik2023provg} across all evaluated datasets and metrics. It also achieves superior precision scores, ranging from 0.95 to 0.98, while maintaining perfect recall (1.00) in most cases. Its area under the curve values ranges from 0.97 to 0.99, demonstrating robust discriminative performance that surpasses the 0.96–0.98 and 0.97–0.98 ranges achieved by MEGR-APT and ProvG-Searcher, respectively.
\par
MEGR-APT’s consideration of temporal sequences and causal relationships in query graphs, coupled with their graph sampling method for threat hunting, contributes to competitive performance; however, it suffers from significant limitations. The system requires substantial preprocessing overhead—approximately 1 h daily for constructing and loading the traceability graph into the RDF database. This preprocessing bottleneck, combined with lower precision scores (MEGR-APT’s 0.93–0.95 vs. \sn{}’s 0.95–0.98), impairs real-time performance and limits practical applicability in time-critical security scenarios. Similarly, while ProvG-Searcher demonstrates consistent performance across datasets, it cannot match the precision advantages offered by \sn{}, particularly evident in the E4 Trace dataset, in which our system achieves a precision of 0.98 compared with that of ProvG-Searcher, which is 0.97.
\begin{table*}[]
\tiny
\centering
\caption{Performance of \sn{} and POIROT. FN denotes the false negative, which occurs when a genuine attack pattern is incorrectly classified as benign. Conversely, FP represents the false positive, where a benign event or data point is mistakenly identified as an attack one. The notation Prec. denotes precision.}
\label{table:eva_metric}
\resizebox{\textwidth}{!}{%
\centering
\begin{tabular}{c|ccc|ccc|ccc} 
\hline
\multirow{2}{*}{\textbf{ATTACK}} & \multicolumn{3}{c|}{\textbf{POIROT}} & \multicolumn{3}{c|}{\textbf{ActMiner without EST}} & \multicolumn{3}{c}{\textbf{ActMiner}} \\ 
\cline{2-10}
 & \textbf{FN/FP} & \textbf{Recall} & \textbf{Prec.} & \textbf{FN/FP} & \textbf{Recall} & \textbf{Prec.} & \textbf{FN/FP} & \textbf{Recall} & \textbf{Prec.} \\ 
\hline
Darpa4 Trace case1 & 19/89 & 99.79 & 99.00 & 21/43 & 99.76 & 99.51 & 0/79 & 100.00 & 99.11 \\
Darpa4 Trace case2 & 58/1005 & 99.62 & 93.89 & 37/532 & 99.76 & 96.62 & 0/781 & 100.00 & 95.18 \\
Darpa4 Trace case3 & 77/82 & 94.58 & 94.24 & 63/41 & 95.51 & 97.03 & 0/65 & 100.00 & 95.39 \\
Darpa4 Trace case4 & 96/1352 & 98.51 & 88.61 & 74/775 & 99.30 & 93.16 & 0/912 & 100.00 & 92.15 \\
Darpa3 FiveDir case1 & 18/254 & 98.52 & 82.49 & 7/88 & 99.42 & 93.19 & 0/142 & 100.00 & 89.40 \\
Darpa3 FiveDir case2 & 49/49 & 96.36 & 96.36 & 32/30 & 97.59 & 97.74 & 0/37 & 100.00 & 97.32 \\
Darpa3 Trace case1 & 77/1146 & 99.47 & 92.62 & 54/412 & 99.62 & 97.20 & 0/613 & 100.00 & 95.91 \\
Darpa3 Trace case2 & 59/1952 & 99.67 & 90.20 & 35/742 & 99.80 & 95.95 & 0/996 & 100.00 & 94.80 \\
Darpa3 Trace case3 & 78/233 & 94.72 & 85.72 & 62/103 & 95.76 & 93.14 & 0/154 & 100.00 & 90.08 \\
Darpa3 Theia & 85/1263 & 99.37 & 91.38 & 63/578 & 99.53 & 95.85 & 0/746 & 100.00 & 94.72 \\ 
\hline
\textbf{Avg FP/FN/Recall/prec.} & 61/742 & 98.06 & 91.45 & 45/\textbf{334} & 98.60 & \textbf{95.94} & \textbf{0}/452 & \textbf{100.00} & 94.41 \\
\hline
\end{tabular}%
}
\begin{tablenotes}
    \centering
    \footnotesize
    \item $\dagger $ \textbf{EST}: Equivalent Semantic Transfer
\end{tablenotes}
\end{table*}

\begin{table}
\centering
\caption{Comparison of ActMiner with GNN-based threat hunting systems across different datasets.}
\label{tab:megr-apt}
\resizebox{\linewidth}{!}{%
\begin{tabular}{|c|c|c|c|c|c|c|} 
\hline
DataSet                    & System         & Precision & Recall & AUC  & TPR  & FPR   \\ 
\hline
\multirow{3}{*}{E4-Trace}  & ActMiner       & 0.98      & 1.00   & 0.99 & 1.00 & 0.02  \\ 
\cline{2-7}
                           & MEGR-APT       & 0.95      & 1.00   & 0.98 & 1.00 & 0.04  \\ 
\cline{2-7}
                           & ProvG-Searcher & 0.97      & 0.99   & 0.98 & 1.00 & 0.03  \\ 
\hline
\multirow{3}{*}{E3-Trace~} & ActMiner       & 0.96      & 1.00   & 0.98 & 1.00 & 0.04  \\ 
\cline{2-7}
                           & MEGR-APT       & 0.95      & 1.00   & 0.98 & 1.00 & 0.03  \\ 
\cline{2-7}
                           & ProvG-Searcher & 0.95      & 1.00   & 0.98 & 1.00 & 0.03  \\ 
\hline
\multirow{3}{*}{E3-Theia~} & ActMiner       & 0.95      & 1.00   & 0.97 & 1.00 & 0.06  \\ 
\cline{2-7}
                           & MEGR-APT       & 0.93      & 1.00   & 0.98 & 1.00 & 0.04  \\ 
\cline{2-7}
                           & ProvG-Searcher & 0.94      & 1.00   & 0.98 & 1.00 & 0.03  \\
\hline
\end{tabular}
}
\end{table}
\subsection{RQ2: How robust is \sn{} against adversarial attacks?}
\label{sec:4.2}
When an attack occurs, the attacker’s behavior pattern may be highly similar or even nearly identical to the normal system behavior in a regular environment. From a technical perspective, attackers can mimic normal processes at the API call level or employ code injection techniques to make their behavior patterns indistinguishable from normal processes at the low-level system log. This poses a challenge for provenance-based threat hunting. However, differentiation can still be achieved by considering richer contextual semantics. 
Goyal et al. \cite{goyal2023sometimes} devise three strategies for adversarial detection of anomaly detection systems based on graph-level granularity. Enlightened by their work, we design experiments to assess \sn{}’s resilience against adversarial attacks.
\par
To assess \sn{}’s resilience against adversarial attacks, we perform adversarial mimicry attacks on provenance-based graph alignment threat hunting system. To evaluate its antiattack capability, using DARPA datasets as a reference, we modify and add attack steps in the provenance graph, primarily considering two scenarios. 
\par
\textbf{Scenario I}: The attacker inserts a large number of invalid attack paths into the actual attack chain, attempting to disrupt and mislead the detection algorithm. Test results show that POIROT exhibited severe missed detections in this scenario, while \sn{} successfully detected all real attack nodes, resisting the attacker’s disruptive attacks. 
\par
\begin{figure}[h]
    \centering
    \includegraphics[width=\linewidth]{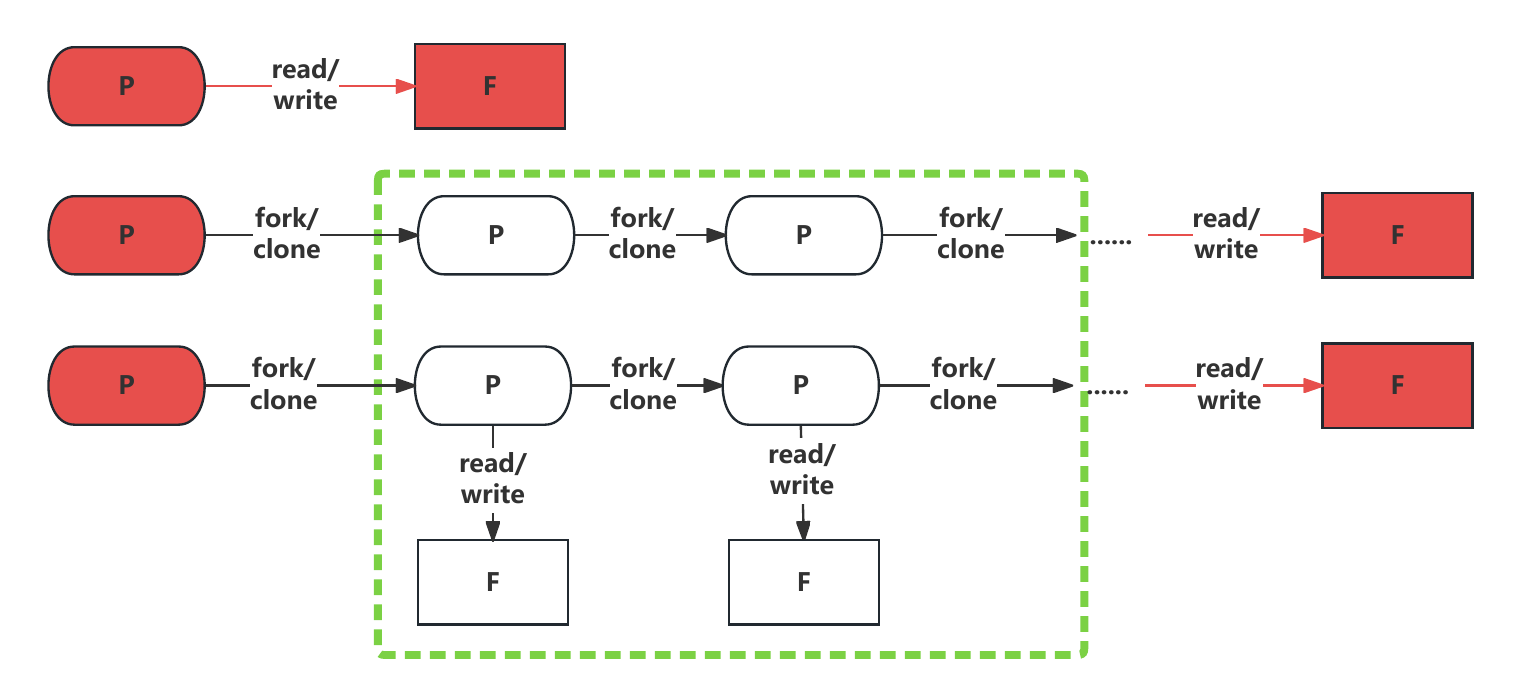}
    \caption{Strategy I. Read/Write operations Insert.}
    \label{fig:1}
\end{figure}
\begin{figure}
    \centering
    \includegraphics[width=\linewidth]{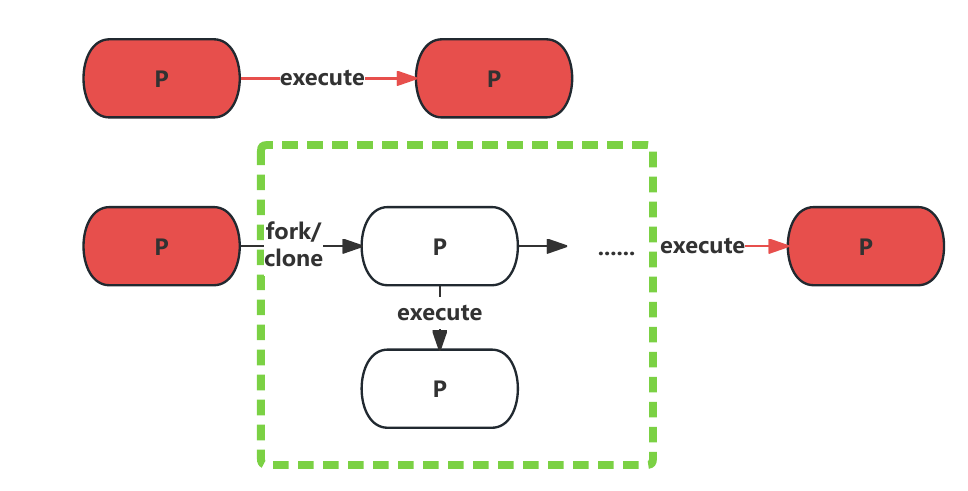}
    \caption{Strategy II. Execute operations Insert.}
    \label{fig:2}
\end{figure}
\begin{figure}[h!t]
    \centering
    \includegraphics[width=\linewidth]{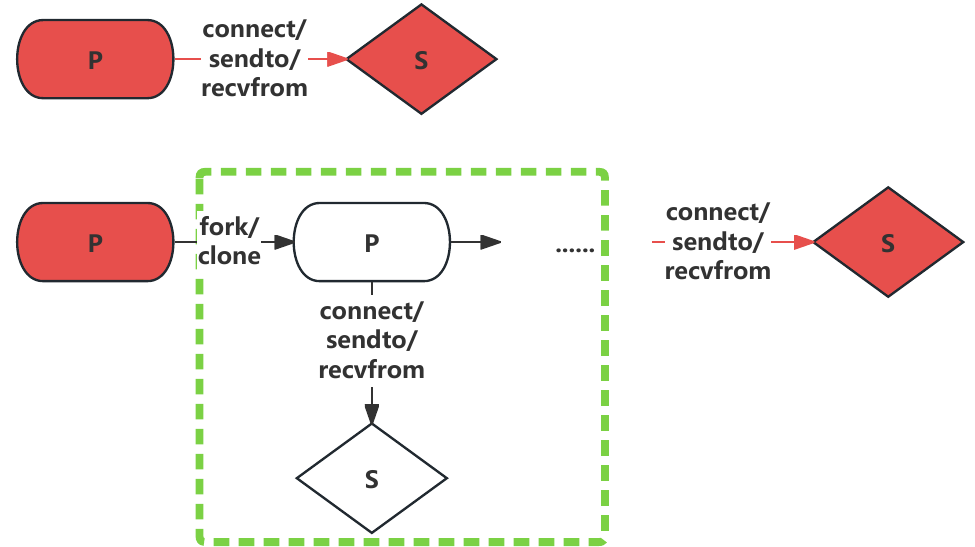}
    \caption{Strategy III. Connect/Send/Recv operations Insert.}
    \label{fig:3}
\end{figure}

\textbf{Scenario II}: The attacker uses normal programs to perform operations similar to attack patterns, attempting to introduce FPs. Test results showed that POIROT exhibits varying degrees of FPs, marking normal processes as attack nodes. In contrast, \sn{} effectively distinguishes the true intent of normal programs through behavior association and intent analysis, avoiding FPs. Based on the above two scenarios, we designed the following three strategies:
\begin{enumerate}
    \item Strategy I: Insert additional unrelated benign process read/write file flows between process and file read/write operations (Figure \ref{fig:1}).
    \item Strategy II: Insert additional benign processes between process and process execution operations (Figure \ref{fig:2}).
    \item Strategy III: Insert additional processes communicating with sockets between processes and socket communication operations (Figure \ref{fig:3}).
\end{enumerate}

The experimental results, as presented in Table \ref{table:anti attack}, depict the percentage of nodes and edges incrementally added to the attack graph using the three aforementioned strategies on the DARPA4 dataset across all attacks. Both nodes and edges are incrementally added at a constant rate.
\par
In terms of recall rate, \sn{} consistently maintains a high average ratio as the proportion of added nodes increases, with a decreasing trend approaching approximately 1\%. This characteristic is determined by the design of \sn{}. In contrast, POIROT exhibits a diminishing recall rate as the proportion increases, with the rate of decline accelerating as the proportions of added nodes increase. This can be observed from the AVG of recall, where the descent shifts from an initial 1.5\% at full scale to around 4\% toward the end of the scale. In terms of precision, as the proportion increases, both \sn{} and POIROT reach their lowest points of 89.33\% and 69.75\%, respectively. Through further analysis, we conclude that \sn{} can detect evasion attacks based on the equivalent semantic approach we have devised.
\par
\textbf{Why can POIROT not defend against adversarial attacks?}
The attacker inserts a large number of irrelevant nodes and connecting edges between the real attack steps, increasing the hop count of a single attack stage to more than three hops. This causes POIROT to fail to associate different stages as a single attack event, resulting in missed detection. POIROT decomposes the graph through bounded branches ($\gamma$) and depth ($\beta$), separating graph neighborhoods from each other. The authors of POIROT consider the optimal depth to be three. Therefore, even if suspicious nodes have a common malicious ancestor, if the hop distance of that behavior from the earliest attack node exceeds $\beta$ hops, the anomaly score obtained by that behavior will be lower than the score of the previous malicious behavior, and that behavior may be mistaken as benign. The attacker can also insert a small number of irrelevant nodes between attack stages, keeping the hop number of the attack chain for each stage within the detection threshold of POIROT to three. Although the inserted nodes themselves are harmless, the edges connecting them to the real attack nodes cause POIROT to erroneously judge them as part of the attack, resulting in FNs.
\par
Through further investigation, we analyze the reasons behind the superior performance of \sn{}: POIROT uses the query graph directly for regular matching of node types and their attribute names and selects appropriate nodes based on path scores, neglecting to consider temporal and causal issues between nodes, resulting in many false alarms.
\par
The robustness of \sn{} results from two reasons. First, attack intent information would be confirmed during the construction of suspicious semantic tree. The asserted behaviors such as fork/clone will not influence the score of the certain attack path, because the multiple hops strategy will consider these nodes to contain the same suspicious semantic. Second, \sn{} does not raise alarms unless nodes exceeding the attack sequence length are detected. This ensures that \sn{} does not generate excessive FPs when attackers intend to conduct multipoint blasting to affect the hunting system while also enhancing its resilience against Scenario I.
\begin{table*}
\scriptsize
\centering
\caption{The performance comparison between the \sn{} and POIROT across four scenarios (i.e., four query graphs) from the E4-Trace dataset, where varying proportions of edges and nodes are added in the provenance graph. Each addition consists of an equal number of three types of nodes/edges. The percentage added represents the percentage of the number of nodes in the provenance graph relative to the attack node.}
\label{table:anti attack}
\resizebox{\textwidth}{!}{%
\centering
\begin{tabular}{|c|cccc|cccc|cccc|}
\hline
\multirow{3}{*}{\textbf{\begin{tabular}[c]{@{}c@{}}Query\\ Graph\end{tabular}}} & \multicolumn{4}{c|}{\textbf{0\%}}                                                              & \multicolumn{4}{c|}{\textbf{25\%}}                                                             & \multicolumn{4}{c|}{\textbf{50\%}}                                                            \\ \cline{2-13} 
                                                                                & \multicolumn{2}{c|}{POIROT}                             & \multicolumn{2}{c|}{ACTMINER}        & \multicolumn{2}{c|}{POIROT}                             & \multicolumn{2}{c|}{ACTMINER}        & \multicolumn{2}{c|}{POIROT}                             & \multicolumn{2}{c|}{ACTMINER}       \\ \cline{2-13} 
                                                                                & \multicolumn{1}{c|}{Precision}    & \multicolumn{1}{c|}{Recall}    & \multicolumn{1}{c|}{Precision}     & Recall     & \multicolumn{1}{c|}{Precision}    & \multicolumn{1}{c|}{Recall}    & \multicolumn{1}{c|}{Precision}     & Recall     & \multicolumn{1}{c|}{Precision}    & \multicolumn{1}{c|}{Recall}    & \multicolumn{1}{c|}{Precision}    & Recall     \\ \hline
case1                                                                           & \multicolumn{1}{c|}{99.00} & \multicolumn{1}{c|}{99.79} & \multicolumn{1}{c|}{100.00} & 100.00 & \multicolumn{1}{c|}{98.34} & \multicolumn{1}{c|}{99.23} & \multicolumn{1}{c|}{100.00} & 100.00 & \multicolumn{1}{c|}{96.24} & \multicolumn{1}{c|}{97.35} & \multicolumn{1}{c|}{99.26} & 100.00 \\ \hline
case2                                                                           & \multicolumn{1}{c|}{93.98} & \multicolumn{1}{c|}{99.62} & \multicolumn{1}{c|}{97.37}  & 100.00 & \multicolumn{1}{c|}{93.28} & \multicolumn{1}{c|}{96.55} & \multicolumn{1}{c|}{97.58}  & 100.00 & \multicolumn{1}{c|}{92.26} & \multicolumn{1}{c|}{94.98} & \multicolumn{1}{c|}{96.33} & 100.00 \\ \hline
case3                                                                           & \multicolumn{1}{c|}{94.24} & \multicolumn{1}{c|}{94.58} & \multicolumn{1}{c|}{98.32}  & 100.00 & \multicolumn{1}{c|}{93.21} & \multicolumn{1}{c|}{94.52} & \multicolumn{1}{c|}{98.42}  & 100.00 & \multicolumn{1}{c|}{91.35} & \multicolumn{1}{c|}{92.25} & \multicolumn{1}{c|}{96.47} & 100.00 \\ \hline
case4                                                                           & \multicolumn{1}{c|}{88.61} & \multicolumn{1}{c|}{98.51} & \multicolumn{1}{c|}{95.08}  & 100.00 & \multicolumn{1}{c|}{87.25} & \multicolumn{1}{c|}{94.31} & \multicolumn{1}{c|}{95.52}  & 100.00 & \multicolumn{1}{c|}{83.21} & \multicolumn{1}{c|}{90.44} & \multicolumn{1}{c|}{94.35} & 100.00 \\ \hline
AVG                                                                             & \multicolumn{1}{c|}{93.96} & \multicolumn{1}{c|}{98.13} & \multicolumn{1}{c|}{97.69}  & 100.00 & \multicolumn{1}{c|}{93.02} & \multicolumn{1}{c|}{96.13} & \multicolumn{1}{c|}{97.88}  & 100.00 & \multicolumn{1}{c|}{90.77} & \multicolumn{1}{c|}{93.76} & \multicolumn{1}{c|}{96.60} & 100.00 \\ \hline
\multirow{3}{*}{\textbf{\begin{tabular}[c]{@{}c@{}}Query\\ Graph\end{tabular}}} & \multicolumn{4}{c|}{\textbf{75\%}}                                                             & \multicolumn{4}{c|}{\textbf{100\%}}                                                            & \multicolumn{4}{c|}{\textbf{125\%}}                                                           \\ \cline{2-13} 
                                                                                & \multicolumn{2}{c|}{POIROT}                             & \multicolumn{2}{c|}{ACTMINER}        & \multicolumn{2}{c|}{POIROT}                             & \multicolumn{2}{c|}{ACTMINER}        & \multicolumn{2}{c|}{POIROT}                             & \multicolumn{2}{c|}{ACTMINER}       \\ \cline{2-13} 
                                                                                & \multicolumn{1}{c|}{Precision}    & \multicolumn{1}{c|}{Recall}    & \multicolumn{1}{c|}{Precision}     & Recall     & \multicolumn{1}{c|}{Precision}    & \multicolumn{1}{c|}{Recall}    & \multicolumn{1}{c|}{Precision}     & Recall     & \multicolumn{1}{c|}{Precision}    & \multicolumn{1}{c|}{Recall}    & \multicolumn{1}{c|}{Precision}    & Recall     \\ \hline
case1                                                                           & \multicolumn{1}{c|}{93.55} & \multicolumn{1}{c|}{95.26} & \multicolumn{1}{c|}{97.56}  & 100.00 & \multicolumn{1}{c|}{91.41} & \multicolumn{1}{c|}{94.32} & \multicolumn{1}{c|}{96.46}  & 99.98  & \multicolumn{1}{c|}{89.74} & \multicolumn{1}{c|}{90.25} & \multicolumn{1}{c|}{95.84} & 99.98  \\ \hline
case2                                                                           & \multicolumn{1}{c|}{90.04} & \multicolumn{1}{c|}{93.52} & \multicolumn{1}{c|}{94.32}  & 100.00 & \multicolumn{1}{c|}{87.43} & \multicolumn{1}{c|}{90.78} & \multicolumn{1}{c|}{94.16}  & 100.00 & \multicolumn{1}{c|}{82.36} & \multicolumn{1}{c|}{87.32} & \multicolumn{1}{c|}{92.52} & 99.98  \\ \hline
case3                                                                           & \multicolumn{1}{c|}{89.21} & \multicolumn{1}{c|}{91.56} & \multicolumn{1}{c|}{95.36}  & 100.00 & \multicolumn{1}{c|}{88.47} & \multicolumn{1}{c|}{90.24} & \multicolumn{1}{c|}{95.13}  & 99.97  & \multicolumn{1}{c|}{86.25} & \multicolumn{1}{c|}{84.33} & \multicolumn{1}{c|}{94.78} & 99.97  \\ \hline
case4                                                                           & \multicolumn{1}{c|}{80.32} & \multicolumn{1}{c|}{87.03} & \multicolumn{1}{c|}{93.42}  & 100.00 & \multicolumn{1}{c|}{77.22} & \multicolumn{1}{c|}{85.41} & \multicolumn{1}{c|}{92.74}  & 100.00 & \multicolumn{1}{c|}{74.32} & \multicolumn{1}{c|}{81.36} & \multicolumn{1}{c|}{92.04} & 99.97  \\ \hline
AVG                                                                             & \multicolumn{1}{c|}{88.28} & \multicolumn{1}{c|}{91.84} & \multicolumn{1}{c|}{95.17}  & 100.00 & \multicolumn{1}{c|}{86.13} & \multicolumn{1}{c|}{90.19} & \multicolumn{1}{c|}{94.62}  & 99.99  & \multicolumn{1}{c|}{83.17} & \multicolumn{1}{c|}{85.82} & \multicolumn{1}{c|}{93.80} & 99.98  \\ \hline
\multirow{3}{*}{\textbf{\begin{tabular}[c]{@{}c@{}}Query\\ Graph\end{tabular}}} & \multicolumn{4}{c|}{\textbf{150\%}}                                                            & \multicolumn{4}{c|}{\textbf{175\%}}                                                            & \multicolumn{4}{c|}{\textbf{200\%}}                                                           \\ \cline{2-13} 
                                                                                & \multicolumn{2}{c|}{POIROT}                             & \multicolumn{2}{c|}{ACTMINER}        & \multicolumn{2}{c|}{POIROT}                             & \multicolumn{2}{c|}{ACTMINER}        & \multicolumn{2}{c|}{POIROT}                             & \multicolumn{2}{c|}{ACTMINER}       \\ \cline{2-13} 
                                                                                & \multicolumn{1}{c|}{Precision}    & \multicolumn{1}{c|}{Recall}    & \multicolumn{1}{c|}{Precision}     & Recall     & \multicolumn{1}{c|}{Precision}    & \multicolumn{1}{c|}{Recall}    & \multicolumn{1}{c|}{Precision}     & Recall     & \multicolumn{1}{c|}{Precision}    & \multicolumn{1}{c|}{Recall}    & \multicolumn{1}{c|}{Precision}    & Recall     \\ \hline
case1                                                                           & \multicolumn{1}{c|}{87.21} & \multicolumn{1}{c|}{88.41} & \multicolumn{1}{c|}{92.36}  & 99.98  & \multicolumn{1}{c|}{82.34} & \multicolumn{1}{c|}{81.64} & \multicolumn{1}{c|}{91.50}  & 99.96  & \multicolumn{1}{c|}{77.28} & \multicolumn{1}{c|}{76.88} & \multicolumn{1}{c|}{90.03} & 98.74  \\ \hline
case2                                                                           & \multicolumn{1}{c|}{77.23} & \multicolumn{1}{c|}{83.42} & \multicolumn{1}{c|}{89.64}  & 99.47  & \multicolumn{1}{c|}{72.14} & \multicolumn{1}{c|}{80.21} & \multicolumn{1}{c|}{88.75}  & 99.48  & \multicolumn{1}{c|}{61.45} & \multicolumn{1}{c|}{76.33} & \multicolumn{1}{c|}{87.25} & 99.52  \\ \hline
case3                                                                           & \multicolumn{1}{c|}{83.52} & \multicolumn{1}{c|}{81.02} & \multicolumn{1}{c|}{93.74}  & 99.86  & \multicolumn{1}{c|}{79.42} & \multicolumn{1}{c|}{77.31} & \multicolumn{1}{c|}{92.36}  & 99.23  & \multicolumn{1}{c|}{75.04} & \multicolumn{1}{c|}{73.98} & \multicolumn{1}{c|}{91.11} & 99.13  \\ \hline
case4                                                                           & \multicolumn{1}{c|}{70.30} & \multicolumn{1}{c|}{79.52} & \multicolumn{1}{c|}{91.14}  & 99.97  & \multicolumn{1}{c|}{68.23} & \multicolumn{1}{c|}{75.21} & \multicolumn{1}{c|}{90.02}  & 99.95  & \multicolumn{1}{c|}{65.23} & \multicolumn{1}{c|}{72.06} & \multicolumn{1}{c|}{88.94} & 99.32  \\ \hline
AVG                                                                             & \multicolumn{1}{c|}{79.57} & \multicolumn{1}{c|}{83.10} & \multicolumn{1}{c|}{91.72}  & 99.82  & \multicolumn{1}{c|}{75.53} & \multicolumn{1}{c|}{78.59} & \multicolumn{1}{c|}{90.66}  & 99.66  & \multicolumn{1}{c|}{69.75} & \multicolumn{1}{c|}{74.81} & \multicolumn{1}{c|}{89.33} & 99.17  \\ \hline
\end{tabular}%
}
\end{table*}
\subsection{RQ3: How important are the components we design for assisting threat hunting?}
\label{section:4.3}
To demonstrate the impact of our system components, particularly the query graph processing (QGP) and EST components on system performance and efficiency, we conduct ablation experiments on them separately. The QGP step is responsible for generating the required query graphs and facilitating the subsequent operations of our system. In comparison, the EST step plays a crucial role in threat hunting module. 
\par
1) \textbf{Efficacy of EST.}
We test the performance of \sn{} without EST using the same settings in RQ1. As presented in Table~\ref{table:eva_metric}, without the EST component, \sn{} achieved an improved FNs by 1.4\% and exhibited FP that is 83.96\% to that of POIROT’s, while it represented an increase compared to \sn{}.
\par
It is attributed to the following reasons: The EST process also inherently involves the causal logic and temporal relationships of the attack to some extent. Therefore, removing this component may introduce a small portion of FPs.
\par
\begin{figure}
    \centering
    \includegraphics[width=1\linewidth]{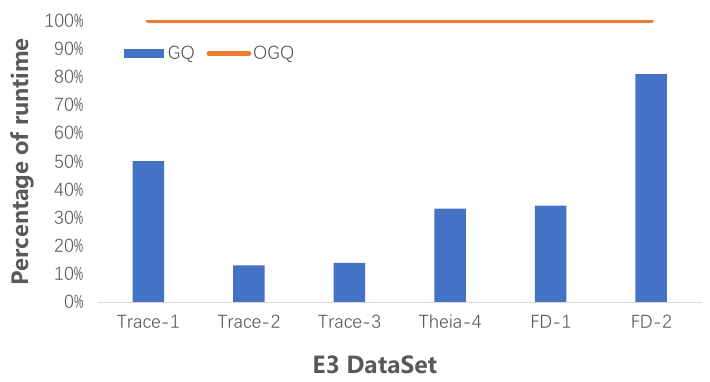}
    \caption{Ratio of running time between GQ and OGQ. OGQ denotes the original query graph, the GQ denotes the query graphs we use in \sn{}.}
    \label{fig:QGP}
\end{figure}

2) \textbf{Effectiveness of QGP.} 
As part of our study, we directly utilize the query graphs extracted by the Extractor and CRUcialG on the same provenance graphs used in RQ1 for threat hunting. The size of the query graph before and after QGP for all situations is presented in Table~\ref{table:QGP}, along with the results and the corresponding analysis. In the QGP process, we use the method of attribute abstraction for nodes, resulting in a reduced number of nodes. Our analysis explores the impact of incorporating QGP on the accuracy and timeliness of the system. We use the Trace dataset from E3 to study the effect of QGP. The result, as depicted in Figure~\ref{fig:QGP}, shows a significant reduction in time consumption when considering QGP. This improvement results from QGP further merging the extracted nodes. Ignoring QGP makes it challenging to efficiently hunt these nodes. However, incorporating QGP allows the system to identify these nodes more effectively, leading to improved threat hunting performance.
\begin{table}
\centering
\caption{The edges and nodes in the query graph, OV and OE denote the number of nodes and edges in the graph before QGP, and V and E denote the number of nodes and edges after processing, respectively.}
\label{table:QGP}
\resizebox{\linewidth}{!}{%
\begin{tabular}{|c|c|c|c|c|c|c|c|c|c|c|} 
\hline
\multirow{2}{*}{Scenario} & \multicolumn{4}{c|}{E4} & \multicolumn{6}{c|}{E3}                \\ 
\cline{2-11}
                          & T-1 & T-2 & T-3 & T-4   & Tr-1 & Tr-2 & Tr-3 & Th-1 & W-1 & W-2  \\ 
\hline
OV                        & 9   & 13  & 16  & 9     & 14   & 10   & 15   & 12   & 13  & 10   \\ 
\hline
OE                        & 8   & 13  & 17  & 8     & 16   & 11   & 14   & 14   & 14  & 11   \\ 
\hline
V                         & 6   & 4   & 6   & 8     & 6    & 8    & 5    & 6    & 5   & 7    \\ 
\hline
E                         & 6   & 3   & 5   & 7     & 5    & 9    & 4    & 5    & 4   & 6    \\
\hline
\end{tabular}%
}
\end{table}
\subsection{RQ4: How efficient is \sn{} compared with the SOTA in terms of runtime overhead?}
\label{section:4.4}
We evaluate how efficiently \sn{} can detect APT attacks in a timely manner by measuring its running time performance. As presented in Table~\ref{tab:time}, POIROT consumes 1.69 times more time than \sn{} under the overall time overhead. This suggests that \sn{} exhibits better timeliness compared with POIROT. However, in some cases, especially the scenario of E3 Trace case1. Through our further investigation, we have discovered a key cause that led to this discrepancy. \sn{} employs a mechanism to off-load data instances spanning over 6 h from memory to the database. While beneficial for optimizing memory management, this storage approach incurs extra overhead when encountering situations that necessitate repeated interactions between the database and memory, resulting in higher time expenditure compared to POIROT. 
\par
Furthermore, the presence of an incremental hunting module allows our system to input data in segments for hunting, rather than having to input all the data at once for each hunt. Specifically, enterprises only need to input the newly added data to the system for each batch (e.g., the data from each day), instead of combining it with data from previous days and performing rescanning. At the same time, we have designed a storage mechanism that stores branches in a database, in which the branch is retrieved from the database when an edge interacting with that node reappears.
\par
\begin{figure}
    \centering
    \includegraphics[width=1\linewidth]{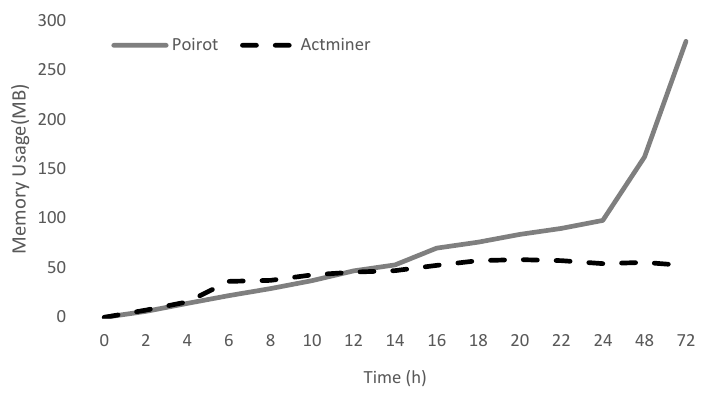}
    \caption{Average Memory consumption in different time states.}
    \label{fig:memory}
\end{figure}

Instead of loading all historical data for each new hunt, only the newly collected information needs to be processed and indexed. As illustrated in Figure \ref{fig:memory}, POIROT’s mean memory requirements gradually increase as more data are collected from hosts over time. However, our incremental hunting module ensures that the rate of memory consumption growth is consistently maintained at a steady pace, primarily determined by our storage mechanism. 
\begin{table}[]
\footnotesize
\centering
\caption{\centering Overhead of \sn{} and POIROT.}
\label{tab:time}
\resizebox{\linewidth}{!}{%
\begin{tabular}{cllll|cllclcll|cllcllcll}
\hline
\multicolumn{5}{c|}{\multirow{2}{*}{\textbf{ATTACK}}} & \multicolumn{8}{c|}{\textbf{POIROT}}                                                                                & \multicolumn{9}{c}{\textbf{ActMiner}}                                                                              \\
\multicolumn{5}{c|}{}                                 & \multicolumn{3}{c}{\textbf{Ti.(s)}} & \multicolumn{2}{c}{\textbf{Mem.(MB)}} & \multicolumn{3}{c|}{\textbf{CPU(\%)}} & \multicolumn{3}{c}{\textbf{Ti.(s)}} & \multicolumn{3}{c}{\textbf{Mem.(MB)}} & \multicolumn{3}{c}{\textbf{CPU(\%)}} \\ \hline
\multicolumn{5}{c|}{E4 Trace case1}               & \multicolumn{3}{c}{2931}            & \multicolumn{2}{c}{65.2}              & \multicolumn{3}{c|}{28.3}             & \multicolumn{3}{c}{5}               & \multicolumn{3}{c}{86.6}              & \multicolumn{3}{c}{25.7}             \\
\multicolumn{5}{c|}{E4-Trace case2}               & \multicolumn{3}{c}{2311}            & \multicolumn{2}{c}{95.6}              & \multicolumn{3}{c|}{28.6}             & \multicolumn{3}{c}{82}              & \multicolumn{3}{c}{118.1}             & \multicolumn{3}{c}{25.7}             \\
\multicolumn{5}{c|}{E4-Trace case3}               & \multicolumn{3}{c}{102}             & \multicolumn{2}{c}{27.8}              & \multicolumn{3}{c|}{26.4}             & \multicolumn{3}{c}{28}              & \multicolumn{3}{c}{35.8}              & \multicolumn{3}{c}{26.0}             \\
\multicolumn{5}{c|}{E4-Trace case4}               & \multicolumn{3}{c}{4284}            & \multicolumn{2}{c}{69.4}              & \multicolumn{3}{c|}{28.3}             & \multicolumn{3}{c}{69}              & \multicolumn{3}{c}{105}               & \multicolumn{3}{c}{25.7}             \\
\multicolumn{5}{c|}{E3-FiveDir case1}             & \multicolumn{3}{c}{5}               & \multicolumn{2}{c}{53.5}              & \multicolumn{3}{c|}{26}               & \multicolumn{3}{c}{26}              & \multicolumn{3}{c}{46.2}              & \multicolumn{3}{c}{21.4}             \\
\multicolumn{5}{c|}{E3-FiveDir case2}             & \multicolumn{3}{c}{73}              & \multicolumn{2}{c}{31.0}              & \multicolumn{3}{c|}{29}               & \multicolumn{3}{c}{6}               & \multicolumn{3}{c}{50.2}              & \multicolumn{3}{c}{23.7}             \\
\multicolumn{5}{c|}{E3-Trace case1}               & \multicolumn{3}{c}{7814}            & \multicolumn{2}{c}{178.1}             & \multicolumn{3}{c|}{26.5}             & \multicolumn{3}{c}{15927}           & \multicolumn{3}{c}{112}               & \multicolumn{3}{c}{27.2}             \\
\multicolumn{5}{c|}{E3-Trace case2}               & \multicolumn{3}{c}{39701}           & \multicolumn{2}{c}{348.8}             & \multicolumn{3}{c|}{25.5}             & \multicolumn{3}{c}{9982}            & \multicolumn{3}{c}{155.4}             & \multicolumn{3}{c}{25.4}             \\
\multicolumn{5}{c|}{E3-Trace case3}               & \multicolumn{3}{c}{62}              & \multicolumn{2}{c}{27.8}              & \multicolumn{3}{c|}{25}               & \multicolumn{3}{c}{149}             & \multicolumn{3}{c}{55.9}              & \multicolumn{3}{c}{25.5}             \\
\multicolumn{5}{c|}{E3-Theia}                     & \multicolumn{3}{c}{3416}            & \multicolumn{2}{c}{280.3}            & \multicolumn{3}{c|}{26.8}             & \multicolumn{3}{c}{9629}            & \multicolumn{3}{c}{105.3}             & \multicolumn{3}{c}{25.0}             \\ \hline
\multicolumn{5}{c|}{\textbf{Avg}}                     & \multicolumn{3}{c}{6070}            & \multicolumn{2}{c}{117.8}           & \multicolumn{3}{c|}{27.1}            & \multicolumn{3}{c}{\textbf{3590}}   & \multicolumn{3}{c}{\textbf{87.1}}    & \multicolumn{3}{c}{\textbf{25.1}}   \\ \hline
\end{tabular}%
}
\end{table}

\subsection{RQ5: How robust is \sn{} in benign dataset?}
To comprehensively evaluate the robustness of \sn{}, we conducted experiments on a sizable benign dataset, introduced in Section \ref{sec:evaluation}. This dataset was collected from multiple users performing typical nonmalicious actions, such as downloading and uploading files, taking backups, browsing the Web, and installing or uninstalling software. Additionally, we included the E3, OpTC dataset, which involves benign activities such as website browsing, checking emails, and SSH log-ins.
\par
We randomly selected three hosts and tested their benign data collected over a period of 1 day from the OpTC dataset and five datasets acquired from our own laboratory.
\par
We applied \sn{} to all benign datasets and searched for the query graphs extracted from the TC reports. Although these logs were attack-free, they shared many nodes and events with our query graphs, such as critical system files and processes related to email clients and text editing tools. As presented in Table \ref{table:benign_results}, despite these similarities, \sn{} successfully demonstrated robustness by generating zero false alerts throughout the experiment.
\begin{table}
\centering
\caption{Experimental results on different benign datasets.}
\label{table:benign_results}
\resizebox{\linewidth}{!}{%
\begin{tabular}{|c|c|c|c|c|} 
\hline
\textbf{DataSet} & \textbf{Test Duration} & \textbf{Platform} & \textbf{Hosts} & \textbf{FP} \\ 
\hline
OpTC & 01d00h00m & Window~ & 3 & 0 \\ 
\hline
E3-FiveDir & 01d03h00m & Window~ & 2 &0 \\
\hline
Our Lab & 01d08h13m & Ubuntu 12.04 x64 & 5 & 0 \\
\hline

\end{tabular}
}
\end{table}

\section{Discussion \& Future Work}
\label{sec:5_discussion}
\textbf{Accuracy of \sn{}.} Due to the accuracy of threat hunting relying on the quality of query graphs, we utilized Extractor to automate the extraction of CTI reports provided by DARPA. We achieved results superior to those of SOTA on well-known APT attack datasets (i.e., E3 and E4). However, due to scarce APT attack samples, we were unable to conduct large-scale experiments and analyses. In practice, accurately extracting query graphs from numerous CTI reports and automatically generating diverse and rational ones can improve threat hunting accuracy in the future.
\par
\textbf{EST of \sn{}.} We acknowledge that the EST module in Actminer introduces additional computational overhead and may generate more FPs in practical deployments, as demonstrated in our experimental evaluation. However, the fundamental advantage of EST lies in its ability to significantly reduce FNs, thereby minimizing the risk of undetected security threats.
In real-world scenarios, security analysts can mitigate the impact of increased FPs through structured alert triage systems that prioritize high-severity incidents and implement automated filtering for common low-risk patterns. The comprehensive coverage of the EST module ensures that subtle attack patterns, which might be missed by traditional detection methods, are captured and flagged for investigation. While this approach requires additional analyst resources for alert processing, the enhanced security posture provided by EST's reduced FN rate offers substantial long-term benefits.
\par
To address data quality concerns, we recommend implementing robust data collection frameworks, such as Spade \cite{gehani2012spade}, Kellect \cite{chen2022kellect}, or eAudit \cite{sekar2023eaudit}, which provide more complete provenance information compared with legacy tools. Future work should focus on developing intelligent alert correlation mechanisms and automated FP suppression techniques to further optimize the practical applicability of the EST module while maintaining its superior detection capabilities.
\par
\textbf{Interpretability of \sn{}.} The outputs of \sn{} inherently provide interpretability, as analysts can derive attack-related information from the query graph reports. However, the query graphs and hunting results are not entirely consistent due to the dynamic nature of APT attacks. We can utilize generative artificial intelligence (e.g., large language model) in the future to automatically transform the attack chains obtained from threat hunting into CTI reports that analysts can clearly understand, aiding in better response.
\par
\textbf{Robustness of \sn{}.} The use of graph processing approaches may lead to the loss of fine-grained details. To validate the robustness of \sn{}, we conducted cross-validation by employing query graphs extracted from diverse CTI reports against distinct provenance graphs. The experimental findings consistently demonstrate that \sn{} effectively avoids generating false alarms across all tested origin query graphs. The identified suspicious nodes do not meet the threshold for triggering an alert.
\par
\textbf{Compression frequency of Actminer.} During the provenance graph preparation process, Actminer only retains the most recent timestamp for consecutive temporal events between the same two entities. By retaining only the most recent timestamps, Actminer significantly reduces the need to store redundant data. In consecutive events of the same entity pair, the latest timestamp usually contains the most relevant semantic information. For example, in network threat intelligence, the latest event may indicate the evolution of an attack or a change in strategy. This design choice leads to a critical limitation: Actminer cannot effectively defend against attacks such as ransomware that rely heavily on event frequency patterns for detection. In practical deployment scenarios, such frequency-dependent attacks require the integration of additional specialized tools, such as Bit Defender \cite{bitdefender} and other complementary security solutions. In future work, it is possible to consider integrating lightweight frequency anomaly detection algorithms as preprocessing steps for provenance graph analysis.
\section{Conclusion}\label{sec:conclusion}

We propose \sn{}, a system that enables the detection of the complete APT attacks chains by applying causality tracking and
increment aligning. It overcomes the issues of low precision, low recall, and low efficiency that existed in previous work. Experimental results show that \sn{} exhibits better detection rates and resilience against adversarial attacks compared to the SOTA.

\section{Acknowledgment}\label{sec:Acknowledgment}
This work is supported in part by the following grants: National Natural Science Foundation of China under Grant No. U22B2028, 62372410, and 62002324. Zhejiang Provincial Natural Science Foundation of China under Grant No. LQ21F020016, and LZ23F020011.  The Fundamental Research Funds for the Provincial Universities of Zhejiang under Grant No. RF-A2023009. Wenzhou Key Scientific and Technological Projects under Grant No. ZG2024007. Wenzhou Basic Scientific Research Projects under Grant No. G2024033. Open Fund Project of Wenzhou Cyber Security Detection and Protection
Technology Research Center under Grant No. WZKF2023001, and
WZKF2023002. "Pioneer" and "Leading Goose" R\&D Program of Zhejiang under Grant No. 2025C01082, and 2025C01013.

\bibliographystyle{unsrt}

\bibliography{ref.bib}

\end{document}